\documentclass[12pt,a4paper]{article}

\usepackage{geometry}
\usepackage{helvet}

\usepackage[utf8]{inputenc}
\usepackage{csquotes}
\usepackage{graphicx}
\usepackage{times}
\usepackage[bottom]{footmisc}	% places footnotes at page bottom
\usepackage{amsmath, amssymb, amsthm, graphicx, color, bm, soul}
\usepackage{amsmath}
\usepackage{authblk}
\usepackage{xcolor}			% to use fonts with colors
\usepackage{caption}
\usepackage{subcaption}
\usepackage{dirtytalk}

\usepackage{titlesec} % change the depth of subsections
\titleformat{\paragraph}
{\normalfont\normalsize\bfseries}{\theparagraph}{1em}{}

\usepackage{physics}

\usepackage{braket} % braket notation
\usepackage[qm]{qcircuit} % quantum circuits
\usepackage{musicography}  % musical symbols
\usepackage{enumerate}
\usepackage{booktabs}
\usepackage{mathtools}
\DeclarePairedDelimiter\ceil{\lceil}{\rceil}
\DeclareUnicodeCharacter{2212}{-}

\makeatletter
\def\mathcolor#1#{\@mathcolor{#1}}
\def\@mathcolor#1#2#3{%
  \protect\leavevmode
  \begingroup
    \color#1{#2}#3%
  \endgroup
}
\makeatother

\usepackage[hyphens]{url}

\title{Quantum Representations of Sound: from mechanical waves to quantum circuits}

\author{Paulo V. Itaboraí}
\author{Eduardo R. Miranda}
\affil{Interdisciplinary Centre for Computer Music Research (ICCMR), 
\authorcr 
University of Plymouth, Plymouth, UK 
\authorcr
\{paulo.itaborai,eduardo.miranda\}@plymouth.ac.uk}

\date{}     
                
\begin{document}

\maketitle

\abstract{By the time of writing, quantum audio still is a very young area of study, even within the quantum signal processing community. This chapter introduces the state of the art in quantum audio and discusses methods for the quantum representation of audio signals. Currently, no quantum representation strategy claims to be the best one for audio applications. Each one presents advantages and disadvantages. It can be argued that future quantum audio representation schemes will make use of multiple strategies aimed at specific applications. The authors also discuss ... .\\ \\ {\bf NOTE:} This is an unedited abridged version of the pre-submission draft of a chapter, with the same title, published in the book \textit{Quantum Computer Music: Foundations, Methods and Advanced Concepts}, by E. R.  Miranda (pp. 223 - 274). Please refer to the version in this book for application examples and a discussion on sound synthesis methods based on quantum audio representation and their potential for developing new types of musical instruments.\\
\url{https://link.springer.com/book/10.1007/978-3-031-13909-3}

\renewcommand{\leftmark}{\sc into}

\section{ Introduction}

 Sounds and images share common grounds. From the point of view of information processing, both are just signals. But obviously, we perceive them differently.

\medskip
Most signal processing methods used for sounds are applicable to images and vice-versa. Their main difference is with respect to dimensionality. For instance, whereas sound is a one-dimensional (1D) signal in the time domain, image is a two-dimensional (2D) one. From a mathematical perspective, the higher the dimension of a signal, the more complex to represent and process it. Therefore, it is logical first to learn how 1D signal representation and processing methods work and then extrapolate to 2D (images), 3D (videos), and beyond. By and large, this is how textbooks on audio and visual signal processing introduce the subject; e.g., \cite{dsp} \cite{book:fourierwavelets}.

\medskip
Thus, from a historical perspective, it would seem reasonable to expect that quantum representations and potential algorithms for sound would have appeared in research avenues before the appearance of those for images and video. Surprisingly, this is not the case. An avid interest in developing quantum algorithms for image processing (primarily for facial recognition and similar applications) produced methods for quantum image representations and processing rapidly, leaving the case of 1D sound behind by almost a decade. This gap is unexpected, given the importance of speech technology to the electronics industry. Even more, if bearing in mind the relatively long-standing desire to develop quantum computing for natural language processing \cite{Meichanetzidis2020}. 

\medskip
The first papers describing how to represent an image on a quantum processor theoretically were published in 2003 \cite{venegas2003storing} and 2005 \cite{venegas2005discrete}. In comparison, the first papers proposing quantum representation methods for sound are from 2015 \cite{wang2016qrda} and 2018 \cite{yan2018flexible}; they are based on a method for images  proposed in 2011 \cite{zhang2013neqr}. Indeed, most quantum signal processing algorithms designed to date are for image applications. The quantum sound community needs to catch up. Hence the motivation for this chapter.

\medskip
The chapter is structured as follows: firstly, section \ref{sec:timeinfo} contains a short introduction that delineates essential aspects and concepts of sound, analogue and digital audio representations that will be used later. It generally explains how the original sound content is transformed from one media to another. The concepts shown in this section will propel us toward the quantum territory with better intuition. This introductory section concludes by giving an initial idea of how quantum information logic will be applied to audio signals.

\medskip
The introduction is followed by a short section (\ref{sec:timeinfo}) that explains how time information is generally encoded in a quantum state. The following note (\ref{subsec:terminology}) identifies some confusion problems present in the nomenclatures used by the literature for quantum audio. It proposes a new naming system to unify and clarify the different strategies used to encode audio information.

\medskip
Then, the two subsequent sessions dive into various definitions and schemes proposed for representing audio in quantum machines using the previously proposed naming system. For instance, section \ref{sec:cbr} explores Coefficient-Based representations, whereas section \ref{sec:sbr} focuses on State-Based ones.

\medskip
Section \ref{sec:summary} summarizes the representations shown in the previous sections. It discusses some obstacles that should be accounted for when considering building and running quantum audio circuits for state-of-the-art quantum hardware.

\medskip
Section \ref{sec:qsp} details some basic quantum signal processing operations and circuit design, such as quantum audio addition and concatenation, as well as sample-by-sample comparison.

\medskip
The last section (\ref{sec:qapp}) punctuates some potential artistic applications in the short and near term. Specifically, there is a case study exploring wavetable synthesis and simple effects that make use of configurable parameters of coefficient-based representations.
\medskip
% {\color{red}{Describe how the chapter is structured ...}}

\medskip

%%%%%%%%%%%%%%%%%%%%
\section{From Mechanical to Quantum}
%%%%%%%%%%%%%%%%%%%%
The main objective of this section is to review how sound is represented in digital computers and electronic devices in general. It will provide the foundations to understand how sound can be represented for quantum computation. 

\medskip
In order to store a given sound in a quantum computer (i.e., by the time of writing), one would need to record the sound as analogue audio and make an analogue-to-digital conversion. Then, the digital audio needs to be translated into a quantum audio representation of some kind (Figure \ref{fig:MTQ}). Let us go through each of these stages in more detail.

\begin{figure}[ht]
    \centering
    \includegraphics[width=480pt]{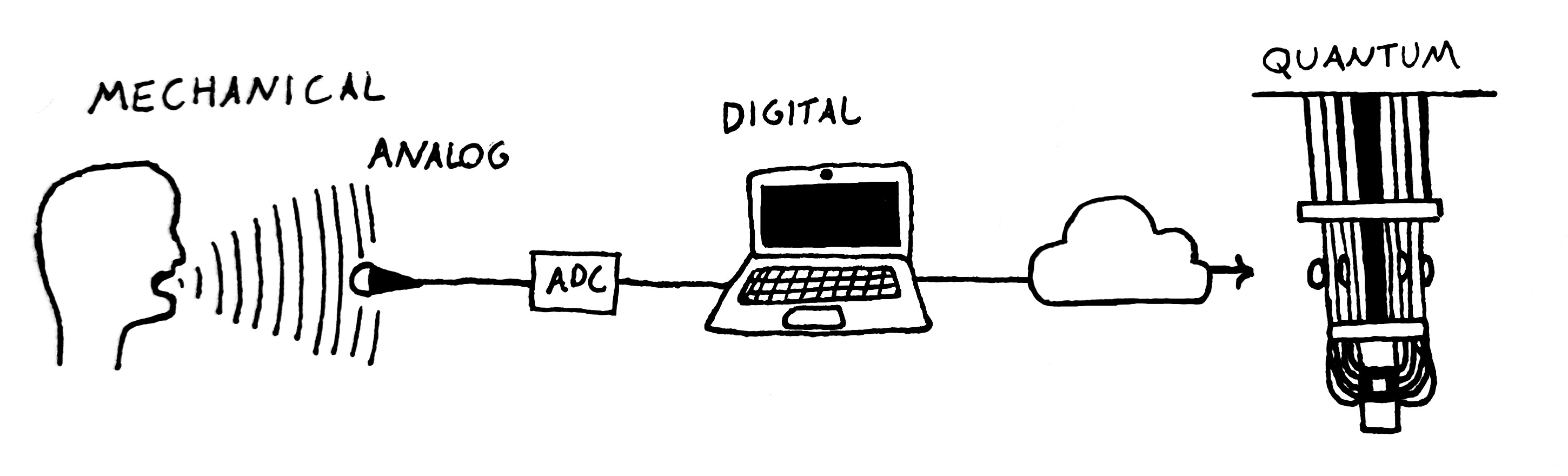}
    \caption{The signal path from mechanical sound to quantum sound.}
    \label{fig:MTQ}
\end{figure}
\vskip1.6cm

\subsection{From Mechanical to Analog}
In the most general sense, waves are physical entities that carry information about disturbances in a particular medium. For instance, electromagnetic waves carry information about disturbances in an electromagnetic field. Sound, however, is a mechanical wave. It can be generated by providing energy to a set of coupled vibrating systems in an acoustic medium, such as air. 

\medskip
Upon reaching a microphone, vibrating air pressure - that is, sound - becomes electric voltage. This is \textit{analog audio}. There are many ways to do this mechanic-to-electric conversion. One of the first techniques ever developed to do this can still be found in dynamic microphones today. 

\medskip
A dynamic microphone has three main elements: a thin diaphragm, a cylindrical magnet and a wire coil. The coil is wrapped around the magnet, but it does not touch it. It can move freely on the cylinder's axis. The diaphragm is coupled to one end of the coil.  So, the vibrating sound will move the diaphragm back and forth. 
As a consequence, the coil will oscillate through the magnet - or, from the perspective of the coil, the magnet will oscillate through it. In an oversimplified way, Faraday's Law teaches us that when a magnet is moving through a conductive coil, it will induce an electric current. This movement will result in a measurable voltage. So, an oscillating coil will induce an oscillating voltage at its terminals. Thus, the mechanical sound has been converted into a varying electric voltage signal which is, by definition, analogue audio.
\subsubsection{Audio Encoding}
\medskip
Analog audio creates a direct connection between mechanical and electrical media. We could say that it \textit{represents sound}. 

\medskip

Analog audio imposes itself as a new technique for propagating and manipulating sound, which was not possible before its invention. For instance, a singer could try to sing louder or even scream, attempting to reach listeners located far away and still not be heard. Alternatively, this singer could convert her voice into audio and then transmit the signal through conductive wires or electromagnetic waves until it reaches the listener. The audio would then be converted back into sound through loudspeakers. The caveat is that transmission, reception and conversion are prone to noise and errors.

\medskip

Attempts to reduce noise and errors during long transmissions (for example, radio transmissions) are what motivated the first audio encoding schemes as effective ways of representing electric audio information. The three most widely used analogue audio encoding schemes are Amplitude Modulation (AM), Frequency Modulation (FM) and Phase Modulation (PM) \cite{modulationbook}. In these cases, the raw analogue signal is encoded by means of a measurable variation of a \textit{signal parameter}, such as the frequency, amplitude (i.e., power), or phase.

%%%%%%%%%%%%%%%%%%%%
\subsection{From Analog to Digital}
%%%%%%%%%%%%%%%%%%%%
\label{subsec:adc}
A digital computer is unable to process a continuous stream of audio, which contains voltage information at all possible points in time. There are infinitely many of them inside an audio signal. Thus, analog information needs to be digitized. This is done by means of an ADC (Analog-to-Digital Converter) device, which converts an analog signal into a digital signal. How is this done? How to turn continuous time and amplitude information into discrete binary codes?

\medskip

This section introduces the notions of \textit{sampling} and \textit{quantization}, which are discretizations of time and amplitude, respectively.

\medskip
Audio sampling is the action of taking snapshots of a continuous signal and storing them (i.e., the values captured by the snapshots) as time-indexed \textit{samples} (Figure \ref{fig:sigSamp}).

\begin{figure}[ht]
    \centering
    \includegraphics[width=450pt]{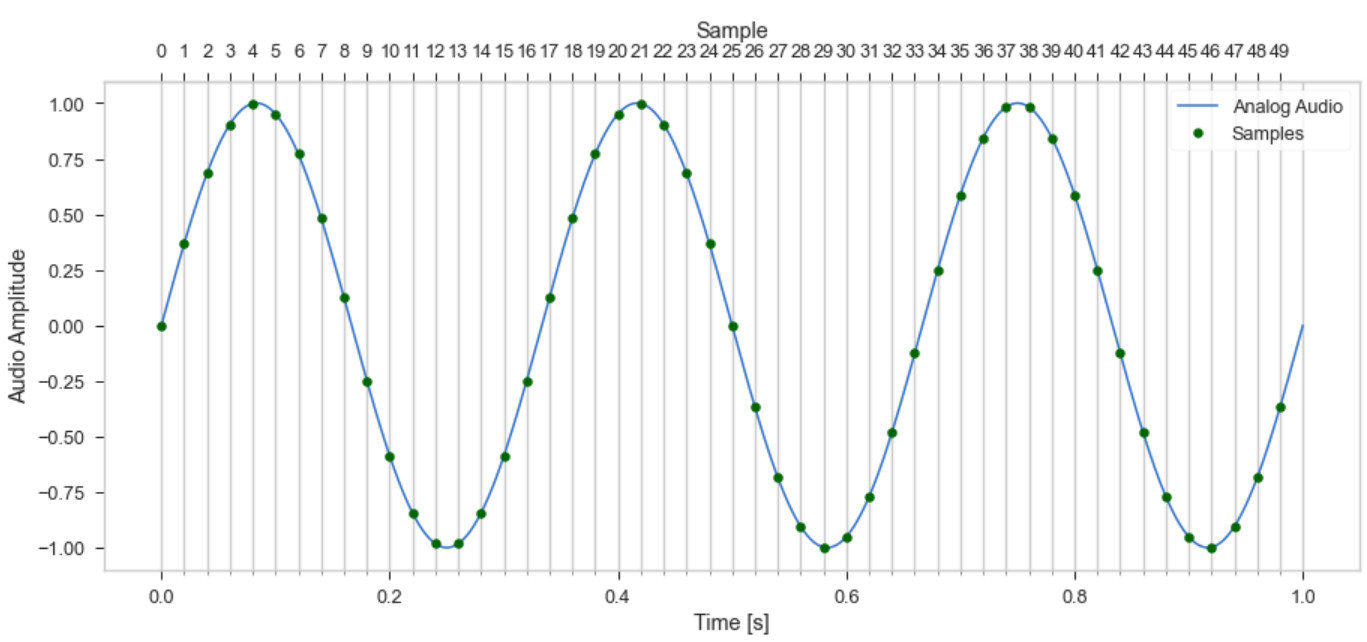}
    \caption{A sampled sine wave.}
    \label{fig:sigSamp}
\end{figure}

\medskip
As shown in Figure \ref{fig:sigSamp}, the snapshots of a signal are conventionally taken in equally spaced time lapses. The speed of lapses is referred to as the \textit{sampling rate} or \textit{sampling frequency}. In other words, the sampling rate establishes how many samples are taken per unit of time. Good quality audio systems use a sampling rate of 44,100 Hz (or 44,100 snapshots per second).
\medskip
The sampling rate has the role of translating an index $k={0, 1, 2, ...}$ into a respective instant in time $t_k$. That is, the sampling rate $S_R$ is the constant responsible for carrying the conversion between the moment of the snapshot (in time units) and an index of time (dimensionless) (Eq. \ref{eq:sampling}). The index $k$ can be represented in binary form and thus stored in a classical computer.

\begin{equation}
    t_k = \frac{k}{S_R}
    \label{eq:sampling}
\end{equation}

\medskip
Now, let us look at the quantization step. Bear in mind that the notion of quantization here is nothing to do with quantum mechanics. Here, quantizing means \textit{to restrict a continuous range to a prescribed set of values}. 

\medskip
Let us examine how the amplitude of an analogue signal, usually represented in the $-1$ to $1$ range, can be quantized into binary numbers (Figure \ref{fig:sigSampQ}).

\begin{figure}[ht]
    \centering
    \includegraphics[width=450pt]{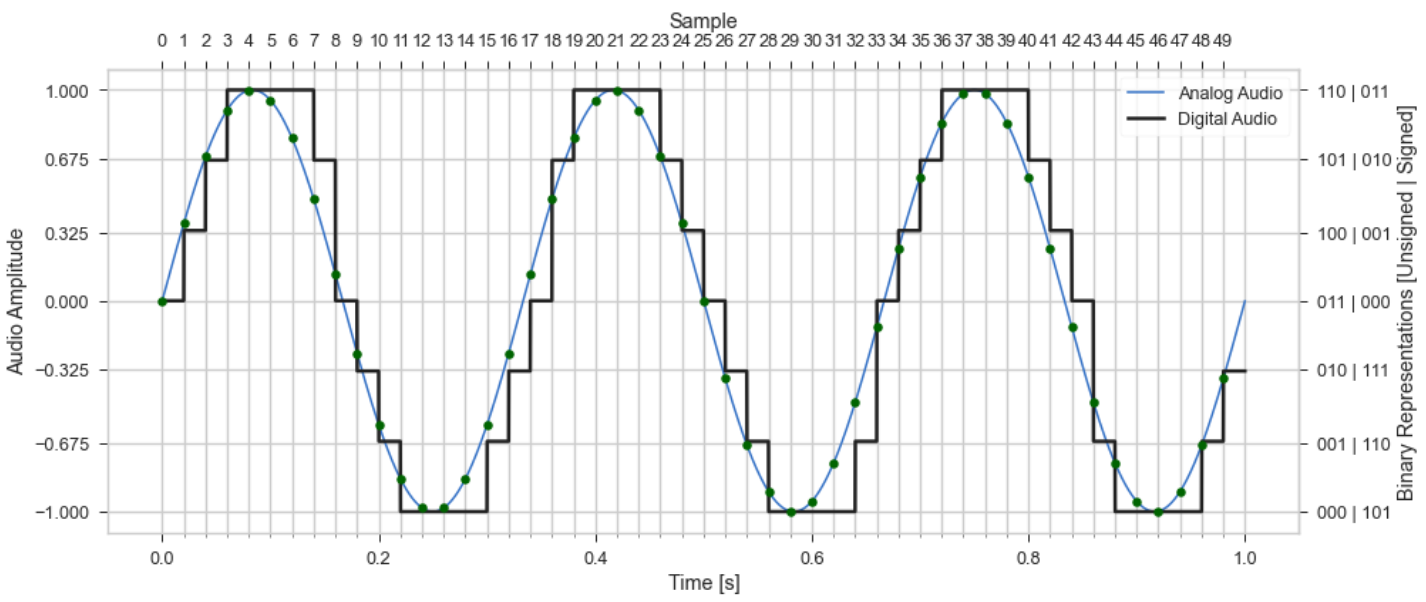}
    \caption{An analog wave and its respective quantization.}
    \label{fig:sigSampQ}
\end{figure}

\medskip
Quantization is, in some ways, similar to sampling, but it operates alongside the vertical axis. The main difference is that the values can be negative. Nevertheless, we can rescale and shift the whole signal to make it fall entirely in the positive domain without any loss of generality. This is a resourceful operation in audio signal processing.

\medskip

The way to shift a signal along a determined axis is by adding or subtracting a defined quantity on all signal values. In our case, we would like to shift the entire signal up by $1$, so we add $1$ to every amplitude value. After doing this, the amplitudes range from $0$ to $2$ instead of  $-1$ and $1$. The range is already positive, but we can further improve this by rescaling the signal to the $0$ to $1$ range; this will be relevant to our quantum audio discussion later. The rescaling is done by multiplication and division, and of course, for our case, we divide the amplitudes by $2$. With this \textit{unitary} range, we can do the quantization scheme more easily. If we have $n$-bit words to store each amplitude, we will perform $2^n$ subdivisions in our range. Then, the voltage values are rounded to fit the closest binary value afforded by the subdivision scheme.

\medskip
Hence, we have successfully digitized our audio by putting them on a grid, as shown in Figure \ref{fig:sigSampQ}.

\subsubsection{Two's Complement Scheme}
\label{subsec:twos}
The quantization scheme presented above is perfectly valid and general for any type of signal in a positive range. However, we have seen that audio signals have negative values as well. 

\medskip
Of course, this is not a problem in itself. We showed above how to shift audio to a positive domain. However, this shifting procedure can often render the audio unsuitable for specific manipulations. A concerning example would be amplitude multiplication. Shifting a value to the strictly positive domain could drastically alter the result of a simple multiplication (as indicated in Eq. \ref{eq:distributive}).

\begin{equation}
    (a_1+shift)(a_2+shift) \neq a_1a_2 + shift
    \label{eq:distributive}
\end{equation}

\medskip
This absence of negative numbers may also prove to be at least inconvenient. Consider, for instance, the task of adding two audio signals.  
The way to add digitized audio waves is by summing their amplitudes, term-by-term. Specifically, imagine an oversimplified digital noise cancellation algorithm. It relies on negative numbers. First, some unwanted noise in an audio signal is received. Then, it generates a copy of the noise with inverted polarity (In other words, all amplitudes are multiplied by $-1$). Finally, the inverted signal is added to the original. Each amplitude is added to a negated version of itself, meaning they cancel each other, and the noise disappears. This phenomenon is nominally called \textit{destructive interference}\footnote{This also appears in the context of the quantum theory, since quantum systems also show wave-like behaviour and thus have wave-like properties such as superposition and interference, among others. But do not be misled. A classical wave and a quantum particle-wave still are fundamentally different, even if they have similar mathematical properties}.
Even though there might be an elaborate way of achieving the desired result with only positive numbers, a less laborious alternative is desirable.

\medskip

A viable solution - which is the one that is generally used for representing audio signals - is to use another interpretation of the binary integers. One that could enable us to represent negative numbers and facilitate binary addition arithmetic. This solution is referred to as the \textit{Two's Complement} scheme.

\medskip
It is important to understand the underlying logic of representing negative numbers in binary, as this will also be useful for quantum audio representations.

\medskip
The main idea of the Two's Complement encoding scheme is to divide the binary words into two sets. Half of them would represent negative numbers and the other half positive ones. This can be made by using the \textit{Most Significant Bit} (i.e., the leftmost bit), or MSB, as a \textit{signing bit}. If the MSB is $0$, then the number is positive. Otherwise, it is negative (Eq. \ref{eq:MSB}).

\begin{equation}
    0010...\longrightarrow\overset{(+\ or -)}{0}\quad \overset{(number)}{010...}
    \label{eq:MSB}
\end{equation}

\medskip
However, just the signing bit by itself is not enough for having a 'good' negative number.
For it to be useful, it needs to comply with binary addition properties.

\medskip
Let us consider a computer with a 4-bit architecture and then use the MSB to indicate the signal of the following 3 bits without applying any other logic. Now, let us try to add $2$ and $-2$ using this new interpretation. Consider the simplified binary addition rules in Eq. \ref{eq:addrules}, and see what happens when we add the numbers (Eq. \ref{eq:addnumbers}).

\begin{equation}
\begin{aligned}
    &0\:+\:0\;=\;0\\
    &0\:+\:1\;=\;1\:+\:0\;=\;1\\
    &1\:+\:1\;=\;10\quad \{\text{\footnotesize{The bit to the left is added by $1$}}\}
     \label{eq:addrules}
\end{aligned}
\end{equation}

\begin{equation}
    2 \rightarrow 0\,010 \, ; \quad -2 \rightarrow 1\,010\;\therefore\;
    0\,010 + 1\,010 = 1\,100 \rightarrow -4 \:(?)
     \label{eq:addnumbers}
\end{equation}

What is shown in Eq. \ref{eq:addnumbers} clearly does not work. But this can be solved. We can apply another condition to our encoding scheme. That is: $x+(-x) = 0$. This condition would interestingly and conveniently change the representation of the negative numbers completely, as shown in Eq. \ref{eq:addsolution}. Now we can verify in Eq. \ref{eq:addperfectly} that the addition works perfectly.

\begin{equation}
\overset{Regular\ Integer}{
\begin{matrix}
    \begin{bmatrix}
    15\\
    14\\
    ...\\
    8
    \end{bmatrix}\\
    \\
    \begin{bmatrix}
    7\\
    ...\\
    1\\
    0
    \end{bmatrix}
\end{matrix}
\quad=\quad
\begin{matrix}
\begin{bmatrix}
        1\ 111\\
    1\ 110\\
    \:...\\
    1\ 000\\
\end{bmatrix}\\
\\
\begin{bmatrix}
    0\ 111\\
    \:...\\
    0\ 001\\
    0\ 000
\end{bmatrix}
\end{matrix}}
\quad\searrow\quad
\overset{Two's\ Complement\ Integer}{
\begin{matrix}
    \begin{bmatrix}
    0\ 111\\
    \:...\\
    0\ 001\\
    0\ 000
\end{bmatrix}\\
\\
\begin{bmatrix}
        1\ 111\\
    1\ 110\\
    \:...\\
    1\ 000\\
\end{bmatrix}
\end{matrix}
\quad=\quad
\begin{matrix}
    \begin{bmatrix}
    7\\
    ...\\
    1\\
    0
    \end{bmatrix}\\
    \\
    \begin{bmatrix}
    -1\\
    -2\\
    ...\\
    -8
    \end{bmatrix}
\end{matrix}}
\label{eq:addsolution}
\end{equation}

\begin{equation}
    2 \rightarrow 0\,010 \, ; \quad -2 \rightarrow 1\,110\;\therefore\;
    0\,010 + 1\,110 = 0\,000 \rightarrow 0
\label{eq:addperfectly}
\end{equation}

%%%%%%%%%%%%%%%%%%%%%%%%%
\subsubsection{Digital Audio as an Array}
%%%%%%%%%%%%%%%%%%%%%%%%%

After being digitized, the audio's binary information is stored in a data structure. The most common structure for digital audio is an array vector, visualized in Figure \ref{fig:array}. The time information becomes an index to a specific position of an array, and the respective amplitude becomes the variable stored in that position. In this way, one can retrieve the amplitudes by having the time index.

\begin{figure}[ht]
    \centering
    \begin{tabular}{|c|c|c|c|c|c|c|c|}
    \hline
        $a_0$ &  $a_1$ & $a_2$ & $a_3$ & $a_4$ & $a_5$ & $a_6$ & $a_7$
    \\
    \hline
    
    \end{tabular}
    
    \begin{tabular}{cccccccc}
    
     \,\footnotesize{$t_0$}\,  & \,\footnotesize{$t_1$}\, & \,\footnotesize{$t_2$}\, & \,\footnotesize{$t_3$}\, & \,\footnotesize{$t_4$}\, & \,\footnotesize{$t_5$}\, & \,\footnotesize{$t_6$}\, & \,\footnotesize{$t_7$}\,
    \end{tabular}
    \caption{Audio array visualization.}
    \label{fig:array}
\end{figure}

\medskip
In order to efficiently represent, store and transmit binary data structure as a stream of audio information, a digital audio encoding scheme is typically used. Similar to the analog case shown above, the digital domain provides some parameters that can be controlled (or modulated) to represent information. For example, we could represent a stream of digital information using \textit{pulses} that are precisely controlled by piezoelectric crystals. The representation that has dominated the music industry is the PCM audio or Pulse Code Modulation. PCM has become a standard in digital audio due to its reliable lossless storage and transmission capacities. Furthermore, it is less prone to noise and errors compared to other pulse representations such as Pulse Amplitude Modulation (PAM), Pulse Position Modulation (PPM), and Pulse Width Modulation (PWM \cite{modulationbook}.

\medskip
We should note at this stage that the two's complement digital audio is used in a wide variety of applications. However, modern 32-bit and 64-bit operating systems, as well as high-level programming languages like Python, further improve the representation of the quantized audio amplitudes by using \textit{floating-point numbers}. A good approximation of Real ($\in \mathbb{R}$) numbers can be stored with floating-point numbers, with high precision and significant decimal figures. Thus, instead of storing an integer representing the middle number of a quantization range, we could add a new pre-processing layer and store a good approximation of the number itself, using the flexibility of floating-point numbers to process the signal in the digital medium. However, this requires a considerable amount of space and computing power. As we will see in the remainder of this chapter, we are still far from similar schemes for quantum computer audio. But imaginable nevertheless.

%%%%%%%%%%%%%%%%%%
\subsection{From Digital to Quantum}
%%%%%%%%%%%%%%%%%%

Gaining an intuitive understanding of quantum audio representation is straightforward, in the sense that Dirac's notation can clearly show how the audio information (time and amplitude) can be translated into quantum states (Figure \ref{fig:digtoq}). Of course, the underlying quantum computing paradigm can be counter-intuitive. However, this intuitive bird's-eye view understanding of the organizational structure of quantum audio gives us an end goal and provides a guiding thread for us to cross the uncertain quantum algorithmic roads. On this path, we shall explore how these representations of quantum audio are prepared and measured, as well as potential near-term and long-term applications.

\begin{figure}[ht]
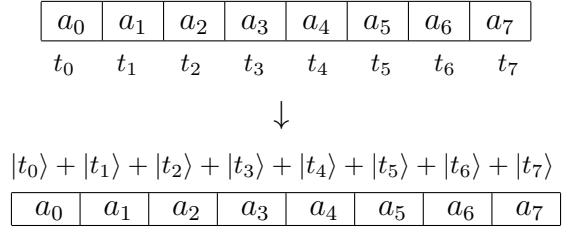

    \centering
    \begin{tabular}{|c|c|c|c|c|c|c|c|}
    \hline
        $a_0$ &  $a_1$ & $a_2$ & $a_3$ & $a_4$ & $a_5$ & $a_6$ & $a_7$
    \\
    \hline

    \end{tabular}
        
    \begin{tabular}{cccccccc}
    
     \,\footnotesize{$t_0$}\,  & \,\footnotesize{$t_1$}\, & \,\footnotesize{$t_2$}\, & \,\footnotesize{$t_3$}\, & \,\footnotesize{$t_4$}\, & \,\footnotesize{$t_5$}\, & \,\footnotesize{$t_6$}\, & \,\footnotesize{$t_7$}\,
    \end{tabular}
    
    \vspace{5pt}
    
    $\downarrow$
    \vspace{7pt}
    
    \footnotesize{$\ket{t_0} + \ket{t_1} + \ket{t_2} + \ket{t_3} + \ket{t_4} + \ket{t_5} + \ket{t_6} + \ket{t_7}$}
    \vspace{4pt}
    
    \begin{tabular}{|c|c|c|c|c|c|c|c|}
    \hline
        \ \normalsize{$a_0$}\ &  \ \normalsize{$a_1$}\ & \ \normalsize{$a_2$}\ & \ \normalsize{$a_3$}\ & \ \normalsize{$a_4$}\ & \ \normalsize{$a_5$}\ & \ \normalsize{$a_6$}\ & \ \normalsize{$a_7$}\ \\
    
    \hline
    \end{tabular}
    \caption{Digital to Quantum}
    \label{fig:digtoq}
\end{figure}

\medskip
It is important to state that quantum audio is still a very young area of study, even under the quantum signal processing umbrella. Therefore many fundamental questions remain open for discussion. There is no audio representation strategy that we could argue to be the best one for audio applications. Each of them presents particular advantages and disadvantages. It can be argued that quantum audio will make use of multiple representations targeting specific applications.

%%%%%%%%%%%%%%%%%%%%%%%%%%%%%%%%%
\section{Preparation and Retrieval of Quantum Audio}
%%%%%%%%%%%%%%%%%%%%%%%%%%%%%%%%%
\label{sec:timeinfo}
The near-term prospects of quantum technology indicate that quantum computers will not replace classical computers in many ways. Rather, they will co-exist, creating hybrid classical-quantum processing environments. This already occurs in current working quantum systems, many of which are accessible via cloud services and interact with classical machines. It is not going to be different for near-term quantum audio applications. These applications rely strongly on classical computers, not only to provide a digital audio file to be prepared but also to post-process quantum circuit measurement results. There are well-known classical programming languages for this part; e.g. Python and MatLab. In the following sections, we will look in detail at quantum audio representations that have been proposed to date. We will examine how quantum audio is prepared, measured, and envisage possible applications. Let us start by introducing the connection between time indexes and quantum superpositions.

%%%%%%%%%%%%%%%%%%%%%
\subsection{Encoding Time Information}

As mentioned previously, a qubit is a 2-state quantum system that can be measured to value either $0$ or $1$. It can also be put in a superposition of states,  written as shown in Eq. \ref{fig:superpos}, where $\abs{\alpha}^2$ and $\abs{\beta}^2$ are the probabilities that a measurement of this state results in $0$ or $1$, respectively. Since the sum of the probabilities of all possible outcomes needs to be $1$, it means that $\abs{\alpha}^2 + \abs{\beta}^2 =1$.

\begin{equation}
    \ket{\Psi} = \alpha\ket{0} + \beta\ket{1}
    \label{fig:superpos}
\end{equation}

\medskip
For 2-qubit states, a similar equation can be written, considering all possible outcomes, as shown in Eq. \ref{eq:2qubits}.

\begin{equation}
    \ket{\Psi} = a\ket{00} + b \ket{01} + c\ket{10} + d\ket{11};\quad \quad \abs{a}^2 + \abs{b}^2 + \abs{c}^2 + \abs{d}^2 = 1
    \label{eq:2qubits}
\end{equation}

\medskip
Since we have few letters in the alphabet to represent many of these probability amplitudes, it might be better to change the notation slightly and use the same letter with a subscript. For instance, a 3-qubit state written this way is shown in Eq. \ref{eq:withsubs}.

\begin{equation}
	\begin{aligned}
     	\ket{\Psi} & = a_{(000)}\ket{000} + a_{(001)}\ket{001} + a_{(010)}\ket{010} + a_{(011)}\ket{011} \\ 
     	& + a_{(100)}\ket{100} + a_{(101)}\ket{101} + a_{(110)}\ket{110} + a_{(111)}\ket{111}
	\end{aligned}
	\label{eq:withsubs}
\end{equation}

\medskip
At this point, we ought to make another helpful change to the notation for improving our intuitiveness. Sometimes, it is convenient (and conventional) to interpret the numbers inside the `kets' not as a sequence of states of individual qubits in a register - but as a classical binary bit string associated with an integer number.

\medskip 
Before we do so, it is imperative to remind us that these zeros and ones inside the 'kets' \textit{are not numbers}. Kets are a simplified notation for writing \textit{vectors}. Therefore, whatever is written inside the ket, is just a conventionalized \textit{label} that refers to a vector: $\ket{label}$. We use numbers as labels to reduce the level of abstraction of those mathematical entities and provide some insight into their use inside a system. In other words, the interpretation above is not changing the state in mathematical terms. Instead, it is just a change of notation. It is essential to have this clear and fresh in our minds to avoid confusion as we introduce the representations next.

\medskip
Thus, by interpreting the qubit states as binary numbers, we can write the same 3-qubit state in Eq. \ref{eq:withsubs} as shown in Eq. \ref{eq:withdec}.

\begin{equation}
    \ket{\Psi}_{3-qubit} = \frac{1}{\sqrt{8}}\big[\ket{0} + \ket{1} + \ket{2} + \ket{3} + \ket{4} + \ket{5} + \ket{6} + \ket{7}\big]
    \label{eq:withdec}
\end{equation}

\medskip
All of the quantum audio representations presented in this text share the same encoding strategy for time information. They use a quantum register to create a superposition of all the possible time indexes associated with each audio sample. This is called a \textit{time register}. Each state will be an index, indicating a position in time, similar to a classical array.

 \medskip
Any information related to this state will encode the respective sample using different strategies (Eq. \ref{eq:ampinf}). For instance, it could use the probability amplitude or another coupled quantum register.

\begin{equation}
    (Amplitude)\ket{t_k}
    \label{eq:ampinf}
\end{equation}

\medskip
Note the necessity of a $2^n$-sized signal for all of the representations. We can use zero padding for using audio with different sizes. We will explore in more detail below how amplitude information is represented in the different schemes.

%%%%%%%%%%%%%%%%%%%%%
\subsection{Note on Nomenclature}
\label{subsec:terminology}
\medskip 
Before we proceed, let us clarify the nomenclature used to refer to the various Quantum Audio Representation (QAR) methods that will be reviewed below. For the sake of clarity and systematization, we propose slight adaptations to the names given in the research papers where these methods were originally introduced.

 \medskip
The representation methods can be grouped into two categories related to how they encode the audio amplitude information and retrieve it back to the classical realm through measurements. The first group contains what is referred to as 'Probabilistic' or 'Coefficient-Based' representations - due to its probabilistic nature when retrieving information classically. The second group include 'Deterministic' or 'State-Based' methods.

 \medskip
As mentioned earlier, quantum audio representation methods are derived from methods developed for representing images rather than sound. The research literature introduced methods such as Quantum Representation of Digital Audio (QRDA), Flexible Representation of Quantum Audio (FRQA), and Quantum Representation of Multichannel Audio (QRMA), which are all State-Based. However, there also are some Coefficient-Based methods for images that can (and will in this chapter) be easily adapted for audio representation.

 \medskip
Also, this chapter intends to reach the signal processing community, and in this endeavour, we will be proposing a new nomenclature system for the already proposed representations. The intention is to correlate, integrate or unify these representations with classic representations in the future. For that, we will use the term `Modulation' as a central spine. This means that we will propose to rename, for example, the Flexible Representation of Quantum Audio (which by itself already has some confusion problems in relation to quantum images\footnote{It would feel logical to induce that the Flexible Representation of Quantum Audio (FRQA) was derived from the Flexible Representation of Quantum Images (FRQI). Unfortunately, this is not the case, since the FRQI is a Probability-Oriented Representation, and the FQRA is a State-Oriented one, derived from the Novel Enhanced Quantum Representation for digital images (NEQR). The choice to name the FQRA as such is unclear and might have been made for historical reasons. The FRQI was one of the pioneering representations for images (like FRQA) and probably the first to be widely studied in the literature with a variety of applications.}) into Quantum State Modulation (QSM), based on the fact that amplitude information is stored in a multi-qubit state. This naming system also paves the way for other coefficient-based audio representations, such as the Quantum Probability Amplitude Modulation (QPAM).

%%%%%%%%%%%%%%%%%%%%%% 
\section{Coefficient-Based Representations}
\label{sec:cbr}
%%%%%%%%%%%%%%%%%%%%%

Suppose that we have some digital audio $A$, with $N=2^n$, $n\in\mathbb{Z}^*$ samples, with each sample quantized to $[-2^{q-1}, -2^{q-1}+1, ..., 2^{q-1}-1]$. That is, $2^q$ possible values.

\medskip
We can induce that the easiest way to represent those samples in a quantum state $\ket{A}$ would be to create a superposition of all of its possible states $t$ (encoding time), weighted by their respective probability amplitudes (encoding the sample value), in a way that resembles an array, but in a quantum superposition (Eq. \ref{eq:cbrep} and Eq. \ref{eq:cbrepex}).

\medskip
We can induce that the easiest way to represent those samples in a quantum state $\ket{A}$ would be first to create a superposition of all of its possible states $t$ (encoding time). Then, each time state would be weighted by their respective probability amplitude (encoding the sample value).

\begin{equation}
\begin{tabular}{|c|}
\hline
$a_n$\\
\hline

\end{tabular}
    \;t_n \longrightarrow  \alpha_i\ket{t_i}
    \label{eq:cbrep}
\end{equation}

\begin{equation}
    \ket{A}_{(3-qubitArray)} = \alpha_{0}\ket{0} + \alpha_{1}\ket{1} + \alpha_{2}\ket{2} + \alpha_{3}\ket{3} + \alpha_{4}\ket{4} + \alpha_{5}\ket{5} + \alpha_{6}\ket{6} + \alpha_{7}\ket{7}
\label{eq:cbrepex}
\end{equation}

%%%%%%%%%%%%%%%%%%%%

\subsection{Quantum Probability Amplitude Modulation: QPAM}

More generally, we could write an arbitrary quantum audio of size $N$, with $n$ (or $\ceil[\big]{\log N}$) qubits and build an encoding scheme where each possible amplitude value of the audio is mapped onto a probability amplitude(Eq.  \ref{eq:QPAM}).

\begin{equation}
    \ket{A_{QPAM}} = \sum_{i=0}^{N -1} \alpha_i\ket{i}
    \label{eq:QPAM}
\end{equation}

\medskip
Quantum computing theory usually presents the upper bound of the sum showing the number of qubits explicitly, like ($2^n -1$). Instead, we chose to use $N=2^n$ as our primary notation, as this will make more sense to the reader with an audio digital signal processing background.

\medskip 
With Eq. \ref{eq:QPAM}, we achieved a simple but still very useful QAR. We refer to this method as Quantum Probability Amplitude Modulation representation of audio, or QPAM. It is a convenient name, as the amplitude information is encoded as probability amplitudes of each quantum state.
%%%%%%%%%%%%%%%%%%%%

\subsubsection{Mapping the Digital Amplitudes to QPAM}

Now, let us examine how to convert a digital signal representation to QPAM representation and vice-versa. 

\medskip
The audio amplitudes $a_n$ are not equal to the probability amplitudes $\alpha_i$ of measuring each sample state. Instead, there is a specific \textit{mapping} between them, which makes this encoding scheme possible. Let us see how this works using the hypothetic snippet of audio depicted in Figure \ref{fig:normal}.

\medskip
In Quantum Mechanics, we can affirm that given an arbitrary qubit $\ket{\phi} = \alpha\ket{0}+\beta\ket{1}$ the probability of measuring the state $\ket{0}$ is $\abs{\alpha}^2$, with the proviso that:

\begin{itemize}
    \item Probabilities are numbers ranging between $0$ and $1$
    \item The sum of probabilities of all possible states should be equal to $1$
\end{itemize}

 \medskip
Digital audio amplitudes in Figure \ref{fig:normal}, however, are numbers ranging between $-1$ and $1$. Their sum does not necessarily add to $1$ at all. So, in order to go from digital to quantum, we need to take the following steps to \textit{normalize} the amplitudes:

\begin{itemize}
    \item Step 1: add $1$ to all amplitudes $a_n$
    \item Step 2: divide the amplitudes by $2$
    \item Step 3: divide again, by the sum of all of the amplitudes
    \item Step 4: take the square root of the result
\end{itemize}

\begin{figure}[ht]
    \centering
    \includegraphics[width=250pt]{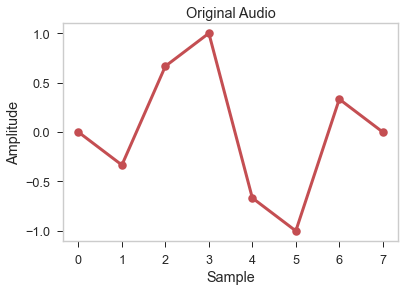}
    \caption{Hypothetic audio.}
    \label{fig:normal}
\end{figure}

 \medskip
Consider the amplitudes of our example listed before the down arrow in Figure \ref{fig:process_norm}. The normalized values are shown after the down arrow. Firstly, we shifted the amplitude values to the positive domain by adding $1$ to every value. At this point, the amplitudes range between $0$ and $2$. Next, we scaled them to fit the range between $0$ and $1$. Then, we summed all their values and divided every value by the result of the sum. At this point, the sum of all amplitudes should be equal to $1$.

\medskip  
The last step is to turn these amplitudes into probability values by taking their square root (Eq. \ref{eq:qa2mapping}).

\begin{figure}[ht]
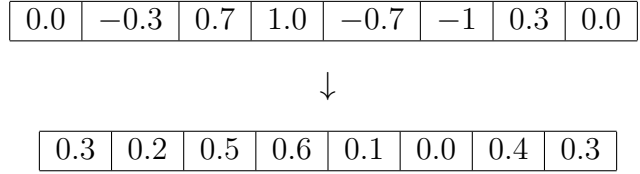

\centering
    \begingroup
        \begin{tabular}{|c|c|c|c|c|c|c|c|}
            \hline
            $0.0$ & $-0.3$ & $0.7$ & $1.0$ & $-0.7$ & $-1$ & $0.3$ & $0.0$ \\
             \hline
        \end{tabular}
        \vspace{10pt} \\
        $\downarrow$ \\
        \vspace{10pt}
        \begin{tabular}{|c|c|c|c|c|c|c|c|}
           \hline
            $0.3$ & $0.2$ & $0.5$ & $0.6$ & $0.1$ & $0.0$ & $0.4$ & $0.3$\\
           \hline
        \end{tabular}
    \endgroup
        
    \caption{Normalization process.}
    \label{fig:process_norm}
\end{figure}

\begin{equation}
    \alpha_i = \frac{1}{\sqrt{g}}\sqrt{\frac{(a_i +1)}{2}}
     ;\qquad g=\sum_k \frac{(a_k +1)}{2}
    \label{eq:qa2mapping}
\end{equation}

%%%%%%%%%%%%%%%%%%

\subsubsection{QPAM Preparation}

The QPAM representation is the most straightforward to prepare. As the information is stored in the probability amplitudes, we just need to initialize the quantum audio at the desired quantum superposition. This can be done by using the probability amplitudes to calculate a unitary gate. Alternatively, one could create a quantum circuit that maps a set of qubits initialized in the $\ket{0000...}$ state into the arbitrary state. Fortunately, we do not need to worry about calculating these matrices and circuits in a more practical sense. Most quantum computing programming tools available nowadays provide commands to initialize qubit states. So, it suffices to assume that any arbitrary superposition of states can be initialized by providing a set of probability amplitudes to the respective command (Figure \ref{fig:inistates}). Consequently, for preparing a QPAM quantum audio, we would only need to convert the digital samples into probability amplitudes.

\begin{figure}[ht]
\centering
\begingroup
    \setlength{\abovedisplayskip}{-1pt}
    \setlength{\belowdisplayskip}{-1pt}
    \begin{equation*}
        \Qcircuit @C=1.0em @R=1.0em @!R {
    	 	\lstick{ {t}_{0} :  } & \multigate{2}{initialize( [\alpha_0, \alpha_1, \alpha_2, \alpha_3, \alpha_4, \alpha_5, \alpha_6, \alpha_7])} & \qw & \qw\\
    	 	\lstick{ {t}_{1} :  } & \ghost{initialize( [\alpha_0, \alpha_1, \alpha_2, \alpha_3, \alpha_4, \alpha_5, \alpha_6, \alpha_7])} & \qw & \qw\\
    	 	\lstick{ {t}_{2} :  } & \ghost{initialize( [\alpha_0, \alpha_1, \alpha_2, \alpha_3, \alpha_4, \alpha_5, \alpha_6, \alpha_7])} & \qw & \qw\\
    	 }
    \end{equation*}
\endgroup
\caption{Initializing qubit states for QPAM.}
\label{fig:inistates}
\end{figure}
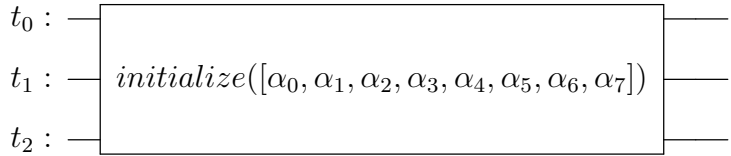

\medskip
Thus, preparing an arbitrary $n$-qubit state usually requires $\order{n}$ simple operations, which implies that QPAM preparation needs $\order{\ceil[big]{\log N}}$ operations.

%%%%%%%%%%%%%%%%%%
\subsubsection{QPAM Retrieval}
 
Retrieving information from the quantum domain means taking measurements from our quantum system. Measuring a quantum system means that only one of the possible state outcomes of that system will be 'materialized' in our measurement apparatus. In other words, once we measure our time registers, the superpositions will be 'destroyed', and only one of the possible time states will be returned.

\medskip
Figure \ref{fig:qpamrep} portrays a hypothetical QPAM representation (i.e., histogram of amplitudes) of the audio shown in Fig \ref{fig:normal}. Furthermore, Figure \ref{fig:qpamretr} shows the respective audio reconstructed by retrieving measurements with an allegedly perfect quantum machine. Let us look into this in more detail.

\begin{figure}[ht]
    \centering
    \includegraphics[width=250pt]{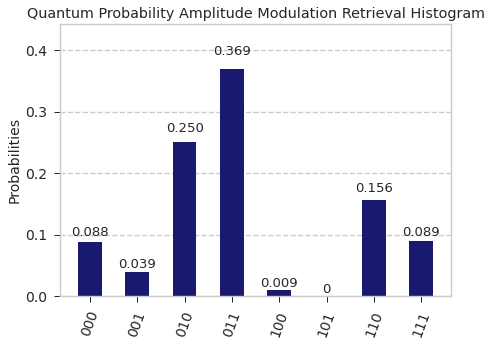}
    \caption{A hypothetical QPAM representation.}
    \label{fig:qpamrep}
\end{figure}

\begin{figure}[ht]
    \centering
     \includegraphics[width=200pt]{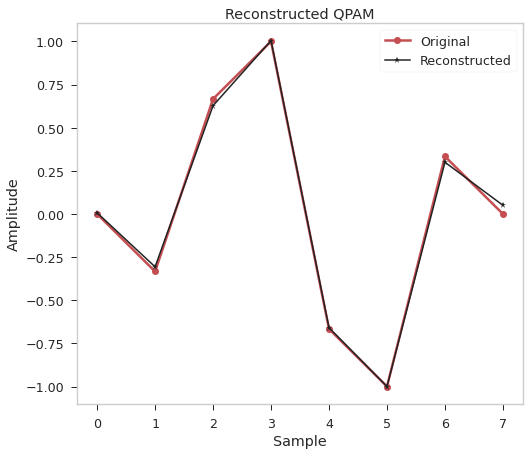}
    \caption{Reconstructed audio from QPAM in Figure \ref{fig:qpamrep}.}
    \label{fig:qpamretr}
\end{figure}

\medskip
The Probabilistic representations have this name precisely because of their retrieval strategy. Retrieving audio requires that we prepare many identical quantum versions of the said audio and make a statistical analysis of the result of all measurements. The analysis will return a histogram of the measurements (e.g., Fig. \ref{fig:qpamrep}), from which we can assess the approximate probability of measuring each state. When considering how the system was prepared in the first place, it indicates that the histogram itself is already a scaled and shifted version of a digital audio. We just need to scale the audio back to its original range, using an inverted version of Eq. \ref{eq:qa2mapping} for $a_i$, shown in Eq. \ref{eq:qa2_mapping_inverted}.

\begin{equation}
    a_i = 2g\abs{\alpha_i}^2 -1
    \label{eq:qa2_mapping_inverted}
\end{equation}

\medskip
There are two critical aspects of Eq. \ref{eq:qa2_mapping_inverted} to be noticed. First, the sum inside the sum is a defined constant value, which is referred to as $g$. Second, the addition of a "$\abs{}$" on $\alpha_i$. We need to be careful with $\alpha_i$ because it is, rigorously, a complex number ($\alpha_i^2 \neq \abs{\alpha_i}^2$). Nevertheless, the valuable information lies in its absolute value squared, which corresponds to the probability $p_i=\abs{\alpha_i}^2$ of $\ket i$ being measured. After completing the measurements, a histogram bin will not return $p_i$, but rather an approximation $\tilde{p}_i$.

\medskip
By replacing equation \ref{eq:qa2_mapping_inverted} with $\tilde{p_i}$, the final post-processing step for retrieving a digital audio is achieved, as shown in Eq. \ref{eq:postproc}.

\begin{equation}
    a_i = 2g\tilde{p}_i -1
    \label{eq:postproc}
\end{equation}

%%%%%%%%
\subsubsection{A caveat}

The constant $g$ in Eq. \ref{eq:qa2mapping} presents an important caveat. The equation contains terms related to digital amplitudes $a_k$. What is the caveat? 

\medskip
Let us consider the preparation stage. The constant $g$ is the denominator, meaning that the sum over the terms $\frac{(a_i +1)}{2}$ cannot be equal to $0$. This condition occurs if all values of $a_i$ were equal to $−1$, which is a perfectly valid audio signal.

\medskip
From a retrieval perspective, there is another consideration. The presence of $g$ makes the normalization of the original input necessary for retrieving a QPAM representation. However, in more general terms, one could consider $g$ to be any arbitrary constant, which would define the overall \textit{amplitude range} of the signal. 

\medskip
This arbitrariness of choice for $g$ can be a problem if not properly assessed. It undesirably changes the amplitude range for the retrieved audio. Let us imagine a thought experiment to illustrate this.

\medskip
Consider a QPAM audio signal with 8 samples. All sample values are equal to $0$. Equation \ref{eq:qa2mapping}, confirms that $g=4$. Then, suppose that a certain quantum algorithm transforms this signal into the measured waveform shown in Fig. \ref{fig:qpamrep}. The signal will be reconstructed using Equation \ref{eq:qa2_mapping_inverted}. Particularly, consider the peak value of the histogram. Applying Eq. \ref{eq:qa2_mapping_inverted} will result in $2g\abs{0.369}^2 -1 = 0.089288$. Moreover, when a low probability sample is picked: $2g\abs{0.09}^2 -1 = -0.999352$. Therefore, the amplitude range is \textit{compressed}.

\medskip
A possible approach to this reconstruction problem may be to use the normalization of the output to calculate a new $g$, as shown in Eq. \ref{eq:newg}.

\begin{equation}
     g=\sum_k \tilde{p}_k
    \label{eq:newg}
\end{equation}

%%%%%%%%%%
\subsection{Single Qubit Probability Amplitude Modulations: SQPAM}

Another coefficient-based quantum audio representation instance can be derived from an image representation scheme known as Flexible Representation of Quantum Images or FRQI \cite{yan2016survey}. We propose calling this method \textit{Single Qubit Probability Amplitude Modulation}, or SQPAM.

\medskip
SQPAM works similarly to QPAM. However, instead of using the raw probability amplitudes, it uses $\ceil{ \log N} +1$ qubits. It improves the logic to encode the samples in the probability amplitudes of one extra, \textit{dedicated} qubit, added as a new register, $\ket{\gamma_i}$, using trigonometric functions. It is a more reliable and editable encoding scheme than QPAM (Eq. \ref{eq:sqpam}).

\begin{equation}
   a_n \longrightarrow \ket{\gamma_i} = \cos{\theta_i}\ket{0} + \sin{\theta_i}\ket{1}
\end{equation}
\begin{equation}
\begin{aligned}
    \ket{A}_{3-qubit} =\frac{1}{\sqrt{8}}\Big[&(\cos{\theta_0}\ket{0} + \sin{\theta_0}\ket{1})\otimes\ket{0} + (\cos{\theta_1}\ket{0} + \sin{\theta_1}\ket{1})\otimes\ket{1} + \\& (\cos{\theta_2}\ket{0} + \sin{\theta_2}\ket{1})\otimes\ket{2} + 
    (\cos{\theta_3}\ket{0} + \sin{\theta_3}\ket{1})\otimes\ket{3} + \\& (\cos{\theta_4}\ket{0} + \sin{\theta_4}\ket{1})\otimes\ket{4} + (\cos{\theta_5}\ket{0} + \sin{\theta_5}\ket{1})\otimes\ket{5} + \\
    &(\cos{\theta_6}\ket{0} + \sin{\theta_6}\ket{1})\otimes\ket{6} + (\cos{\theta_7}\ket{0} + \sin{\theta_7}\ket{1})\otimes\ket{7}\Big]
\end{aligned}
\end{equation}
Generalizing for a N-sized audio, he have:
\begin{equation}
    \ket{A_{SQPAM}}  = \frac{1}{\sqrt{N}}\sum_{i=0}^{N -1} (\cos{\theta_i}\ket{0} + \sin{\theta_i}\ket{1})\otimes\ket{i} 
    \label{eq:sqpam}
\end{equation}

\subsubsection{Mapping the SQPAM Audio Amplitudes}

Similarly to the previous representation, there are some pre-processing steps for mapping the amplitudes to a normalized scope. Trigonometric functions are often used to represent probability amplitudes since their absolute value ranges from 0 to 1, and the relation between the cosine and sine functions satisfies the normalization requirement perfectly\footnote{ 
$\quad\cos{(\theta)}^2+ \sin(\theta)^2 = 1$}.

\medskip
Essentially, this describes a simple change of variables, where $\alpha_i = \cos\theta_i$ and $\beta_i = \sin\theta_i$. Notice how only real numbers are being considered, analogous to QPAM.

\medskip
The introduction of an extra qubit $\ket{\gamma_i}$ significantly improves the representation in terms of both encoding and retrieval of the audio. The first improvement is that the encoding scheme does not rely on the entirety of the audio size content (the total number of states on the superposition and their relative amplitudes) anymore. It is encoded locally in the new qubit and therefore could be independently manipulated, using \textit{rotation matrices} as gates. This opens a new range of possibilities.

\medskip
The function that maps the audio amplitudes $a_i$ into angles $\theta_i$ is displayed in eq. \ref{eq:sqpam_mapping}. This mapping can also be conceived as the following set of instructions:
\begin{itemize}
    \item Step 1: Add 1 to all amplitudes $a_n$.
    \item Step 2: Divide the amplitudes by 2.
    \item Step 3: Take the square root.
    \item Step 4: Evaluate the \textit{Inverse Sine}  of the result (The \textit{Inverse Cosine} is also applicable, as long as the same convention is followed throughout the implementation).
\end{itemize}

\begin{equation}
    \theta_i = \sin^{-1}\left({\sqrt{\frac{a_i +1}{2}}}\right)
    \label{eq:sqpam_mapping}
\end{equation}

\medskip
The mapped array of angles will enable the preparation of the SQPAM quantum audio state.

\subsubsection{Preparation}
The SQPAM state preparation requires $n$ qubits for encoding time (like QPAM) and 1 qubit for the amplitude ($\gamma_i$). They can all be initialized in the $\ket{0}$ state, as denoted in Eq. \ref{eq:0state_notation}.

\begin{equation}
    \ket{000...} \equiv \ket0\otimes\ket0\otimes\ket0\otimes... \equiv \ket0^{\otimes n+1} \equiv \underset{(\gamma_i)}{\ket0}\otimes\ket{0}^{\otimes n}
    \label{eq:0state_notation}
\end{equation}

\medskip
The qubits of the time register are put in a balanced superposition by applying Hadamard gates in each one of them (Figure \ref{fig:sqpam_hadamard}). The resulting quantum state is written in Eq. \ref{eq:sqpam_hadamard}. 

\begin{figure}[ht]
    \centering
    \begingroup
    \setlength{\abovedisplayskip}{-1pt}
    \setlength{\belowdisplayskip}{-1pt}
    \begin{equation*}
        \Qcircuit @C=1.0em @R=0.0em @!R {
    	 	\lstick{ {\gamma} :  } & \qw & \qw & \qw\\
    	 	\lstick{ {t}_{0} :  } & \gate{H} & \qw & \qw\\
    	 	\lstick{ {t}_{1} :  } & \gate{H} & \qw & \qw\\
    	 	\lstick{ {t}_{2} :  } & \gate{H} & \qw & \qw\\
    	 }
    \end{equation*}
    % $$
    \endgroup
    \caption{Circuit segment with Hadamard gates applied to a 3-qubit time register.}
    \label{fig:sqpam_hadamard}
\end{figure}
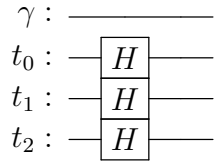

\begin{equation}
    \ket{\Psi} = \frac{1}{\sqrt{N}}\Big[\ket0\otimes\ket0 \;+\; \ket0\otimes\ket1 \;+\; \ket0\otimes\ket2 \;+ ... +\; \ket0\otimes\ket{N-1}\Big]
    \label{eq:sqpam_hadamard}
\end{equation}

\medskip
At this stage, we should discuss the particular strategy for transforming the amplitude qubit into the desired form ($\ket0 \rightarrow \ket{\gamma_i}$). There is a perfect quantum gate for the task, called $R_y$. It applies rotations in the $y$ axis of a Bloch Sphere (Eq. \ref{eq:ry2th}\footnote{$R_y(2\theta)$ has the same form of a 2D rotation matrix, found in many Linear Algebra textbooks. The main difference is that the angle theta rotates twice as fast in a Bloch Sphere compared to a regular Euclidean space.}).  
\begin{equation}
    R_y(2\theta) =
    \begin{pmatrix}
    \cos\theta & -\sin\theta \\
    \sin\theta & \cos\theta
    \end{pmatrix}
    \label{eq:ry2th}
\end{equation}

\medskip

\medskip
Let us take a look at what happens when we apply $R_y(2\theta)$ to the state $\ket0$. Reminding that quantum states represent vectors, this will resolve in a matrix-vector multiplication (Eq. \ref{eq:ry2th_times_0}).

\begin{equation}
    R_y(2\theta_i)\ket0 =     
    \begin{pmatrix}
    \cos\theta_i & -\sin\theta_i \\
    \sin\theta_i & \cos\theta_i
    \end{pmatrix}  
    \begin{pmatrix}
        1 \\ 0
    \end{pmatrix} =
    \begin{pmatrix}
        \cos\theta_i \\ \sin\theta_i
    \end{pmatrix}=
    \cos\theta_i\ket0 +\sin\theta_i\ket1 = \ket{\gamma_i}
    \label{eq:ry2th_times_0}
\end{equation}

% \medskip
% The existence of a rotation gate like this makes the state preparation way more convenient and transparent.

\medskip
Lastly, there should be a way to, somehow, \textit{loop} through the different indexes of the time register for applying a particular $R_y(2\theta_i)$ to them. As a result, the angles $\ket{\gamma_i}$ would be entangled to their respective states $\ket i$. How to access and correlate specific states if they are in a superposition? 

\medskip
The desired correlations can be achieved by controlling our value setting gate $R_y$ using a \textit{multi-controlled} version of $R_y$. Let us explore the effect of controlled gates on a quantum state.

\medskip
Imagine a 1-qubit length audio example. Assume that a superposition was created in the time register by following the first preparation step. The quantum state would be described by Eq. \ref{eq:mcg_phi0}.
% \begin{equation*}
%     \Qcircuit @C=1.0em @R=0.0em @!R {
% 	 	\lstick{ {\gamma} :  } & \qw & \qw & \qw\\
% 	 	\lstick{ {t}_{0} :  } & \gate{H} & \qw & \qw\\
% 	 }
% \end{equation*}

\begin{equation}
    \ket{\phi_0} =\frac{1}{\sqrt{2}}\Big[\ket{0}\ket0 +\ket{0}\ket1\Big]
    \label{eq:mcg_phi0}
\end{equation}

\medskip
Then, a controlled $R_y(2\theta)$ gate is applied. By analyzing the circuit in Fig. \ref{fig:mcg_phi1}, we could conclude that the $R_y$ gate will only be applied if the time qubit is 1. Eq. \ref{eq:mcg_phi1} shows the resulting state.

\begin{figure}[ht]
    \centering
    \begingroup
    \setlength{\abovedisplayskip}{-1pt}
    \setlength{\belowdisplayskip}{-1pt}
\begin{equation*}
    \Qcircuit @C=1.0em @R=0.0em @!R {
	 	\lstick{ {\gamma}_{0} :  } & \qw & \gate{R_y(2\theta)} & \qw & \qw\\
	 	\lstick{ {t}_{0} :  } & \gate{H} & \ctrl{-1} & \qw & \qw\\
	 }
\end{equation*}
    \endgroup
    \caption{Application of a controlled $R_y(2\theta$)}
    \label{fig:mcg_phi1}
\end{figure}
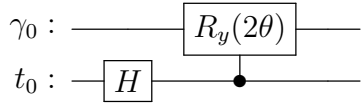

\begin{equation}
    \ket{\phi_1} = cR_{y[t_0]}\ket{\phi_0} =\frac{1}{\sqrt{2}}\Big[\ket{0}\ket0 +R_y(2\theta)\ket{0}\ket1\Big]
    \label{eq:mcg_phi1}
\end{equation}

\medskip
Now, let us place an $X$ gate before and after $R_y$ (Fig. \ref{fig:mcg_phi2}). What is the effect? If the time qubit happens to be in the $\ket0$ state, it will be flipped to $\ket1$. Consequently, it triggers the control condition, which leads to Eq. \ref{eq:mcg_phi2}. Then, it is flipped back to $\ket0$. Notice that it would not trigger the controlled gate otherwise.
\begin{figure}[ht]
    \centering
    \begingroup
    \setlength{\abovedisplayskip}{-1pt}
    \setlength{\belowdisplayskip}{-1pt}
\begin{equation*}
    \Qcircuit @C=1.0em @R=0.0em @!R {
	 	\lstick{ {\gamma}_{0} :  } & \qw & \qw & \gate{R_y(2\theta)} & \qw & \qw & \qw\\
	 	\lstick{ {t}_{0} :  } & \gate{H} & \gate{X} & \ctrl{-1} & \gate{X} & \qw & \qw\\
	 }
	\qquad\equiv\qquad\qquad     
	\Qcircuit @C=1.0em @R=0.0em @!R {
	 	\lstick{ {\gamma}_{0} :  } & \qw & \gate{R_y(2\theta)} & \qw & \qw\\
	 	\lstick{ {t}_{0} :  } & \gate{H} & \ctrlo{-1} & \qw & \qw\\
	 }
\end{equation*}
    \endgroup
    \caption{Inverting the control condition using two X gates}
    \label{fig:mcg_phi2}
\end{figure}
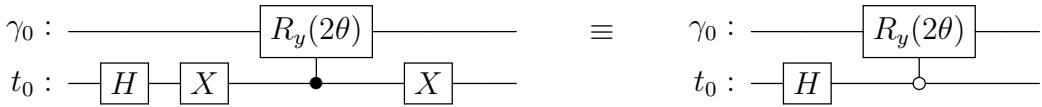

\begin{equation}
    \ket{\tilde{\phi_!}} = (I\otimes X)cR_{y[t_0]}(I\otimes X)\ket{\phi_0} =\frac{1}{\sqrt{2}}\Big[R_y(2\theta)\ket{0}\ket0 +\ket{0}\ket1\Big]
    \label{eq:mcg_phi2}
\end{equation}

\medskip
The $X$ gates create a strategy for effectively switching the control condition, triggering when the qubit is in the $\ket0$ state instead of $\ket1$. This is denoted by a white circle at the control position, as shown on the right side of Fig. \ref{fig:mcg_phi2}.

\medskip

For two-qubit systems, the same strategy can be used to access particular indexes. In this case, there are two control conditions to verify simultaneously. For instance, the address for the third sample - $\theta_2$ - is $\ket{10}$. Hence, the control condition on $t_0$ (the rightmost bit) is "$\ket0$", whereas $t_1$ is "$\ket1$". Equation \ref{eq:2q_mcg} demonstrates that $R_y(2\theta_2)$ is only being applied to the third position.

By using four 2-controlled $R_y(2\theta_i)$ gates, each amplitude can be addressed and stored accordingly (Fig. \ref{fig:2q_mcg}).

\begin{equation}
\begin{matrix}
    \ket{\phi_0} = \frac{1}{2}\Big[\ket0\ket{00} + \ket0\ket{01} +\ket0\ket{10}+\ket0\ket{11}\Big] \therefore\\ \\
    \ket{\phi_{\theta_2}}= (I\otimes X\otimes I) (ccR_{y[t_0t_1]}(\theta_2)) (I\otimes X\otimes I) \ket{\phi_0} \\ \\ = \frac{1}{2}\Big[\ket0\ket{00} + \ket0\ket{01} +R_y(\theta_2)\ket0\ket{10}+\ket0\ket{11}\Big]
\end{matrix}
    \label{eq:2q_mcg}
\end{equation}

\medskip
By generalizing this strategy, we arrive at an algorithm for the second preparation step of the SQPAM audio.
For each time state,
\begin{itemize}
    \item Have a binary version of the state's label stored in a classical variable.
    \item If a bit in a given position of the string is zero, apply an $X$ gate at the respective qubit.
    \item Apply the $R_y(2\theta)$ gate.
    \item Apply another $X$ gate, following the second instruction.
\end{itemize}

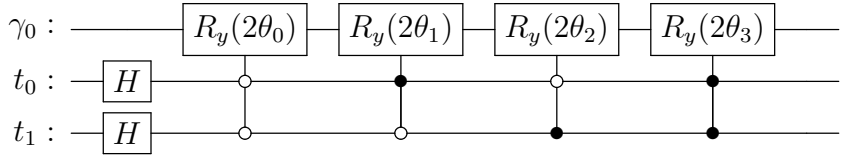
\begin{figure}[ht]
    \centering
    \begingroup
    \setlength{\abovedisplayskip}{-1pt}
    \setlength{\belowdisplayskip}{-1pt}
\begin{equation*}
    \Qcircuit @C=1.0em @R=0.0em @!R {
	 	\lstick{ {\gamma}_{0} :  } & \qw & \gate{R_y(2\theta_0)} & \gate{R_y(2\theta_1)} & \gate{R_y(2\theta_2)} & \gate{R_y(2\theta_3)} & \qw & \qw\\
	 	\lstick{ {t}_{0} :  } & \gate{H} & \ctrlo{-1} & \ctrl{1} & \ctrlo{-1} & \ctrl{1} & \qw & \qw\\
	 	\lstick{ {t}_{1} :  } & \gate{H} & \ctrlo{-1} & \ctrlo{-2} & \ctrl{-1} & \ctrl{-2} & \qw & \qw\\
	 }
\end{equation*}
    \endgroup
    \caption{Preparation of a SQPAM representation using Value-Setting operations with $R_y$ gates}
    \label{fig:2q_mcg}
\end{figure}

\medskip
Notice how this strategy depends on classical numbers. Depending on the programming language of the framework, this can be easily done while constructing the circuit, using a classical 'for loop'.

\medskip
This preparation strategy (using multi-controlled gates) can be called \textit{Value-Setting Operation}. This operation will be applicable to other representations as well.

\medskip
This preparation process applies $\ceil{log N}$ Hadamard gates, and $N$ $N$-controlled $R_y(2\theta)$ gates. Since the $N$-controlled $R_y(2\theta)$ have $\order{N}$ of basic operations, this means that $\order{N^2}$ operations are needed for preparing a SQPAM audio.

\subsubsection{Retrieval}
\medskip
Figure \ref{fig:sqpamRetrieval1} shows what to expect of a SQPAM retrieval histogram. Each sample is encoded in complementary sine and cosine pairs embedded in $\ket{\gamma_i}$ probabilities (Eq. \ref{eq:sqpam_prob}). Hence, 16 bins instead of 8. 

In Equation \ref{eq:sqpam_mapping}, there is a $\arcsin$ function. Therefore, the audio's profile appears if the sine components ($p_{\gamma_i}(\ket1)$) are picked. 

Then, it suffices to use Eq. \ref{eq:sqpam_rec} to reconstruct the signal\footnote{Consider using the cosine term ($p_{\gamma_i}(\ket0)$) in the nominator of Eq. \ref{eq:sqpam_rec}. What would happen? The complementarity of the trigonometric functions would result in a reconstructed audio with \textit{inverted polarity}.}. Figure \ref{fig:sqpamRetrieval2} shows a signal reconstruction using the histogram in Fig \ref{fig:sqpamRetrieval1} and Equation \ref{eq:sqpam_rec}.

\begin{equation}
    p_{\gamma_i}(\ket0) = (\cos(\theta_i))^2 \qquad ; \qquad p_{\gamma_i}(\ket1) = (\sin(\theta_i))^2 \, 
    \label{eq:sqpam_prob}
\end{equation}

\begin{equation}
    a_i = \frac{2p_{\gamma_i}(\ket1)}{p_{\gamma_i}(\ket0)+ p_{\gamma_i}(\ket1)} -1
    \label{eq:sqpam_rec}
\end{equation}

\begin{figure}[t]%{0.55\textwidth}
    \centering
    \includegraphics[width=290pt]{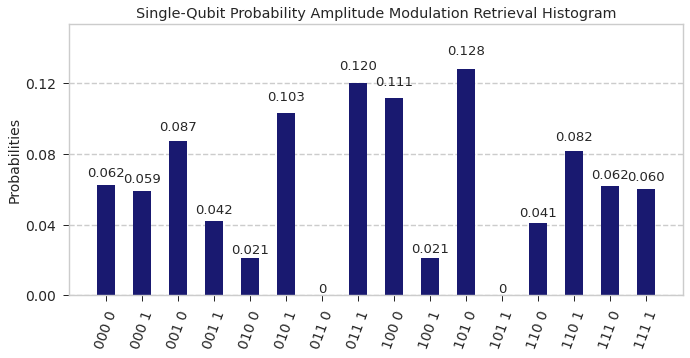}
    \caption{Hypothetical SQPAM representation}
    \label{fig:sqpamRetrieval1}
\end{figure}
    \hspace{30pt}
\begin{figure}[t]%{0.3\textwidth}
    \centering
    \includegraphics[width=180pt]{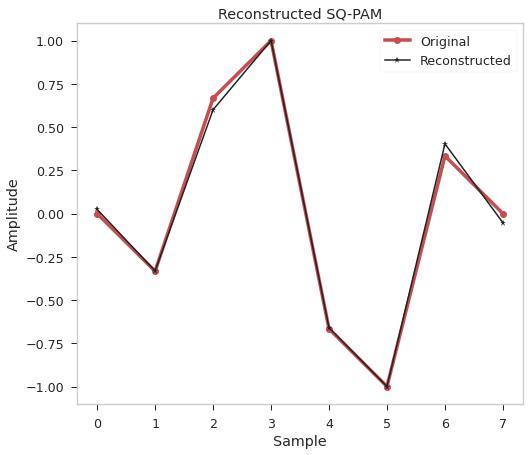}
    \caption{Reconstructed audio from SQPAM in Figure \ref{fig:sqpamRetrieval1}}
    \label{fig:sqpamRetrieval2}
\end{figure}

\section{State-Oriented Representations}
\label{sec:sbr}
In the previous section, the quantum audio representation schemes stored the time domain into a superposition of multi-qubit states and the amplitudes into probability amplitudes. In this section, the main difference is that the audio amplitudes will also be stored in a multi-qubit state. This new quantum register will produce a set of time-amplitude pairs of \textit{entangled information} when put together with the time register. Put another way, if a given time sample $t_i$ is measured, there will be a correlated state that will provide the respective amplitude once measured. 

\medskip

It can be seen as an evolutionary branch of the previous representations (regarding the quantum image representation timeline). For instance, they can \textit{potentially} be manipulated in more detail and used in a wider range of applications. Of course, this needs to be discussed on a case-by-case basis. There is no way to know which representation type is more adequate for different audio and musical applications yet. This will be further discussed in the next section. 

\medskip

The consequence of these "finer" capabilities is that they might increase the amount of space required to represent and manipulate the audio. In other words, there is a trade-off between the number of qubits required to store the quantum audio and the amount of known potential applications. 

\medskip

In addition, another historical motivation to develop the following representation strategies is related to their preparation complexity (e.g. the number of instructions needed for encoding). Furthermore, the logic of storing the audio amplitudes as binary quantum states align with the quantized digital amplitudes of section \ref{subsec:adc}.

\subsection{QSM and uQSM}

The two following quantum audio representations were derived from the Novel Enhanced Quantum Representation of Images, or NEQR \cite{zhang2013neqr}. They were the first proposed quantum representations of audio, namely, QRDA (Quantum Representation for Digital Audio) and FRQA (Flexible Representation of Quantum Audio). Both are identical from a notation standpoint, but there is a slight difference in the interpretation of the qubit states. Respectively, they are read as either \textit{unsigned} or \textit{signed} integers.

\medskip
The QRDA \cite{wang2016qrda} was the first quantum audio representation proposed in the literature. It was responsible for indicating the path towards its preparation and retrieval techniques, bringing many of the primary foundations of a State-Oriented Representation. In section \ref{subsec:adc}, we discussed how the analogue amplitudes were converted into digital. The samples were approximated to the closest binary integer of the quantization scheme. The same quantized amplitudes will be used in this section to prepare multi-qubit states.

\medskip
The QRDA has a dedicated quantum register that encodes the sample using strictly positive integers (directly inherited from the NEQR Image that encoded Gray colour information ranging from 0 to 1). The FRQA, or Flexible Representation of Quantum Audio, on the other hand, uses the Two's Complement scheme to interpret the amplitudes (Section \ref{subsec:twos}). 

The signed integer makes audio mixing extremely convenient since only a regular bit-wise adder algorithm is needed. This indicates that the FRQA's interpretation of the states is more suitable for quantum audio than QRDA.

\medskip
Apart from the variable class interpretation, the logics of preparation and retrieval are identical. These similarities suggest merging their names onto our proposed naming system: Quantum State Modulation, or QSM. Whenever a specific application requires unsigned integers explicitly (QRDA), we could indicate that with a lower case `u' before the acronym (uQSM). This terminology can be extended to fit other variable types in the future, like fixed-point (Sec. \ref{subsec:fpqsm}) or floating-point numbers.

\subsection{QSM}

Consider a digital audio file with $8$ samples and amplitude values quantized in a $3$-bit depth two's complement scheme. Let us assume that the seventh sample of the signal was represented by the following integer bit string: "$001$". In possession of this string, we somehow prepare an analogous 3-qubit string on a dedicated amplitude register by initializing it in the ground state and then flipping the rightmost qubit. This amplitude state can be later entangled with a respective state of the time register, in this case, $t_7$, as described in Eq. \ref{eq:qsm_entangle}.

\begin{equation}
    (a_7, t_7) \equiv (001,110) \longrightarrow \ket{001}\ket{110} \equiv \ket{a_7}\otimes\ket{t_7}
    \label{eq:qsm_entangle}
\end{equation}

\medskip
Applying this structure for all of the samples would result in the superposition shown in equations \ref{eq:qsm1} and \ref{eq:qsm2}. More generally, the QSM could be written as in Eq. \ref{eq:qsm_3}. 

\begin{equation}
\begin{aligned}
      \overset{binary\, notation}{\ket{A_{QSM}} }= \frac{1}{\sqrt{8}}\Big[& \ket{000}\otimes\ket{000}
      \;+\; \ket{111}\otimes\ket{001}
      \;+\; \ket{010}\otimes\ket{010}
      \;+\; \\&\ket{011}\otimes\ket{011}
      \;+\; \ket{110}\otimes\ket{100}
      \;+\; \ket{101}\otimes\ket{101}
      \;+\; \\&\ket{001}\otimes\ket{110}
      \;+\; \ket{000}\otimes\ket{111}\Big]
\end{aligned}
\label{eq:qsm1}
\end{equation}
\begin{equation}
\begin{aligned}
      \overset{decimal \,notation}{\ket{A_{QRDA}} }= \frac{1}{\sqrt{8}}\Big[& \ket{0}\otimes\ket{0}
      \;+\; \ket{-1}\otimes\ket{1}
      \;+\; \ket{2}\otimes\ket{2}
      \;+\; \ket{3}\otimes\ket{3}
      \;+\; \\&\ket{-2}\otimes\ket{4}
      \;+\; \ket{-3}\otimes\ket{5}
      \;+\; \ket{1}\otimes\ket{6}
      \;+\; \ket{0}\otimes\ket{7}\Big]
\end{aligned}
\label{eq:qsm2}
\end{equation}
% In a more generalized way:
\begin{equation}
\begin{aligned}
    \ket{A} = \frac{1}{\sqrt{N}} \sum_{i=0}^{N-1} \ket{S_i}\otimes\ket{i}
\end{aligned}
 \label{eq:qsm_3}
\end{equation}

\medskip
The time register uses $n$ qubits (audio length), and the amplitude register needs $q$ qubits (qubit-depth).

\medskip
As previously mentioned, this strategy has some foreseeable advantages. The multi-qubit register for amplitude directly relates to the digital audio. This similarity suggests that some quantum algorithms could be developed to apply the same operations as known digital signal processing. From there, developers could explore if the quantum media would improve, extend or optimize some of them. This "intuition" layer contributes to the instant popularity of the State-Based representations compared to the less explored Coefficient-Based ones.

\subsubsection{Preparation}
\label{subsec:qsmprep}
The same strategy mentioned in the SQPAM case will be used here: a Value-Setting Operation. In this case, instead of using $R_y(2\theta_i)$, the operation will use multi-controlled CNOT gates for flipping the qubit states of the amplitude register. As a result, the algorithm will access and entangle specific amplitude states with their respective time states.

\medskip
Value-Setting Operation: For each time state,
\begin{itemize}
    \item Step 1: Have a binary form of the time state's label stored in a classical variable.
    \item Step 2: Have the binary form of the quantized audio sample at the same index
    \item Step 3: If a bit in a given position of the time variable is zero, apply an X gate at the respective qubit. Verify all bits.
    \item Step 4: If a bit in a given position of the audio sample is 1, apply a multi-controlled CNOT gate. Verify all bits.
    \item Step 5: Repeat the third instruction to reset the control condition.
\end{itemize}

\medskip
Figures \ref{fig:qsm_testaudio} and \ref{cir:qsm_preparation} exemplify a preparation circuit for an audio with 8 samples and 3-bit depth
\begin{figure}[ht]
    \centering
    \includegraphics[width=250pt]{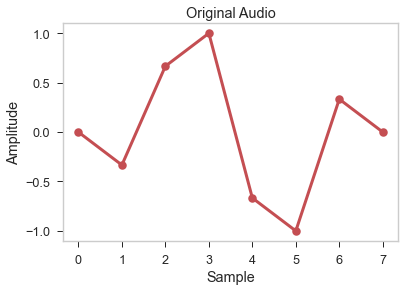}
    \caption{Example audio used to prepare the QSM circuit}
    \label{fig:qsm_testaudio}
\end{figure}

\begin{figure}[ht]
    \centering
    \begingroup
    \setlength{\abovedisplayskip}{-1pt}
    \setlength{\belowdisplayskip}{-1pt}
    \begin{equation*}
    % \vspace{40pt}
    \Qcircuit @C=1.0em @R=0.0em @!R {
	 	\lstick{ {a}_{0} :  } & \qw \barrier[0em]{5} &   \qw & \targ & \qw & \qw  \barrier[0em]{5} & \qw & \qw \barrier[0em]{5} & \qw & \targ & \qw \barrier[0em]{5} & \qw & \qw & \qw \barrier[0em]{5} & \qw & \targ & \qw \barrier[0em]{5} & \qw & \targ & \qw \barrier[0em]{5} & \qw & \qw & \qw\\
	 	\lstick{ {a}_{1} :  } & \qw &   \qw & \qw & \targ & \qw  & \qw & \targ & \qw & \qw & \targ & \qw & \targ & \qw & \qw & \qw & \qw & \qw & \qw & \qw & \qw & \qw   & \qw\\
	 	\lstick{ {a}_{2} :  } & \qw &   \qw & \qw & \qw & \targ &   \qw & \qw & \qw  & \qw & \qw & \qw & \qw & \targ  & \qw & \qw & \targ & \qw & \qw & \qw & \qw & \qw  & \qw\\
	 	\lstick{ {t}_{0} :  } & \gate{H} & \qw &   \ctrl{-3} & \ctrl{-2} & \ctrl{-1} & \qw & \ctrlo{-2} & \qw & \ctrl{-3} & \ctrl{-2} &  \qw  & \ctrlo{-2} & \ctrlo{-1}   & \qw & \ctrl{-3} & \ctrl{-1} & \qw   & \ctrlo{-3}  & \qw & \qw & \qw & \qw \\
	 	\lstick{ {t}_{1} :  } & \gate{H} & \qw & \ctrlo{-1} & \ctrlo{-1} & \ctrlo{-1} & \qw & \ctrl{-1} & \qw & \ctrl{-1} & \ctrl{-1} & \qw & \ctrlo{-1} & \ctrlo{-1} & \qw & \ctrlo{-1} & \ctrlo{-1}  & \qw & \ctrl{-1} & \qw & \qw & \qw & \qw\\
	 	\lstick{ {t}_{2} :  } & \gate{H} & \qw &   \ctrlo{-1} & \ctrlo{-1} & \ctrlo{-1}  & \qw  & \ctrlo{-1} & \qw & \ctrlo{-1} & \ctrlo{-1} & \qw & \ctrl{-1} & \ctrl{-1} & \qw & \ctrl{-1} & \ctrl{-1} & \qw & \ctrl{-1} & \qw & \qw & \qw & \qw\\	 }
    \end{equation*}
    \endgroup
    \caption{Preparation of a QSM representation using Value-Setting operations}
    \label{cir:qsm_preparation}
\end{figure}
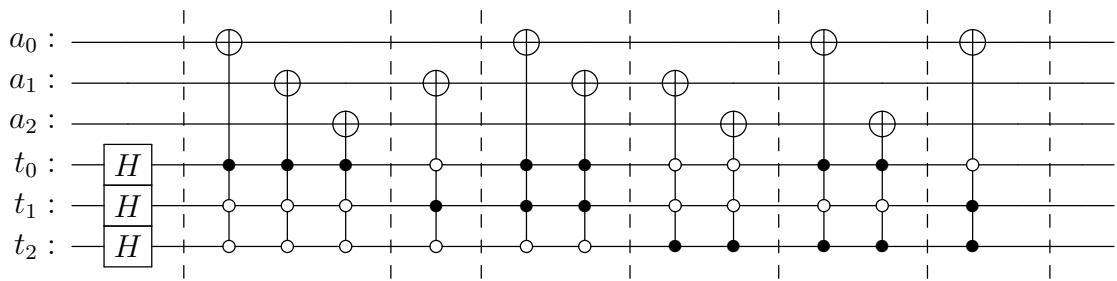

\medskip
Compared with SQPAM ($R_y(2\theta_i)$), the preparing circuit for the audio in fig. \ref{fig:qsm_testaudio} will apply $N$ value-setting operations, demanding $\order{q + \log N}$ basic instructions. In conclusion, the preparation circuit for the QSM is exponentially faster, requiring $\order{qN\log N}$ simple operations.

\subsubsection{Retrieval}
The interesting part of retrieving a state-based quantum audio is that it can be seen as \textit{deterministic}. It guarantees a perfect reconstruction of the digital audio. This is assured since the sample is represented as a binary state.

\medskip
In the coefficient-based version, the samples were statistical quantities that needed to be approximated through many measurements and statistical analysis. There will always be an intrinsic statistical error whenever the amount of measurements is finite. It is a trade-off between lying an error inside some tolerance range (depending on musical, psychoacoustic, perceptual factors) and time consumption (mainly the number of times a circuit will be prepared and measured) \footnote{In his paper about generative quantum images using a coefficient based image representation, James Wootton \cite{wootton2020procedural} states that he measured the $n$-qubit image state $4^n$ times before considering it was a good approximation for his application. This number can be much higher for reliable retrieval, and it scales exponentially.}.

\medskip
In the case of state retrieval, we would only need to measure all valid time-amplitude pairs to retrieve the audio. Each time state is entangled to an amplitude state. So this theoretically assures that a given time state would certainly measure its correlated amplitude; there is no other combined state possible. 

\medskip
That being the case, we could in principle measure each time state just once, enabling the retrieval of all respective amplitudes.
Unfortunately, it is potentially impossible to devise a method that could guarantee only one measurement per sample based on a quantum time state in a superposition. In other words, \textit{controlled measurements} (only measuring a quantum state based on another quantum state) are \textit{non-unitary} instructions and cannot be written as a set of quantum gates. \footnote{We can build circuits that use the result of measurements for controlling quantum gates (due to the \textit{deferred measurement principle}). But only unitary instructions.}.

\begin{figure}[ht]
    \centering
    \includegraphics[width=210pt]{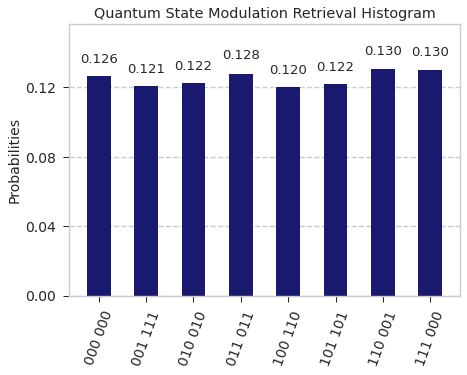}
    \caption{Hypothetical QSM histogram}
    \label{fig:qsm_retrieval1}
\end{figure}
\hfill
\begin{figure}[ht]
    \centering
    \includegraphics[width=180pt]{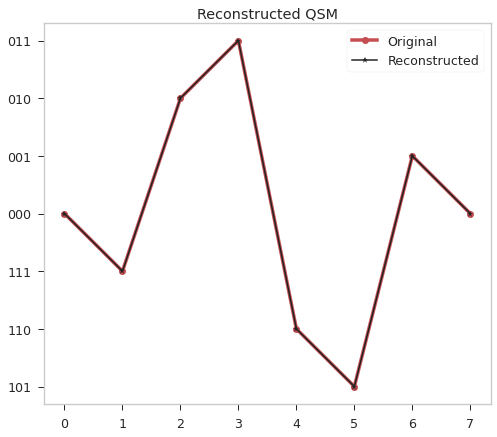}
    \caption{Reconstructed audio from QSM in Figure \ref{fig:qsm_retrieval1}}
    \label{fig:qsm_retrieval2}
\end{figure}

\medskip
With that said, the strategy for retrieving the QSM will be similar in many ways to the previous ones: preparing several copies, making measurements, and analyzing a histogram. Enough measurements are needed to assure a sufficient probability of having all possible states measured. Then, the histogram is analyzed and the \textit{valid} measured states are parsed into digital amplitudes and time indexes. 
This analysis is also a good strategy for dealing with quantum error in real-life quantum processors. Making enough measurements to separate noise from a signal. In an ideal situation, a state-modulated representation will have a balanced superposition of the possible states (states containing the audio amplitudes and their time index), meaning they should be equally likely to be measured (Fig. \ref{fig:qsm_retrieval1}). Moreover, the measured states of the amplitude register are decoded into the correspondent digital amplitudes using the Two's Complement scheme (Fig. \ref{fig:qsm_retrieval2}).

\medskip

The QSM framework is powerful due to its relatability with digital representations. It looks like PCM audio in a superposition. It promptly awakes the idea of comparing the computational efficiency of known classical DSP algorithms (of course, even if quantum parallelism helps with computational complexity, we need to consider the preparation and retrieval time consumption as well).

\subsection{Fixed Point QSM}
\label{subsec:fpqsm}

\medskip
Preparation complexity and quantum states interpreted as integers made QSM (FRQA) well known. Nevertheless, it raises the question: why stop at integers? We could easily interpret the qubit registers as fixed (or floating) point numbers and use the same state structure of the QSM. A fixed point, QSM-based representation was formalized by Li et al. \cite{Li2018digitalsig} and was named QRDS (Quantum Representation of Digital Signals). For the terminology purposes of this text, this representation could be named fixed-point QSM or fpQSM.

\medskip
It essentially divides the $q$ qubits used to represent integers into three parts: one sign bit, $m$ integer bits and $q-m-1$ fractional bits\footnote{the $q-m-1$ was adapted to our notation. In the original notation, the text uses $n+1$ qubits instead of $q$}, as shown in Eq. \ref{eq:fpqsm}.

\begin{equation}
    \ket{A} = \frac{1}{\sqrt{N}} \sum_{i=0}^{N-1} \ket{fixed_i}\otimes\ket{i} = \frac{1}{\sqrt{N}} \sum_{i=0}^{N-1} \ket{x_i^m,x_i^{m-1}\cdots x_i^0\ .\ x_i^{-1}\cdots x_i^{q-m-1}}\otimes\ket{i}
    \label{eq:fpqsm}
\end{equation}

\subsection{Multichannel Audio}
\label{sec:multichannel}
\medskip
The QRMA (Quantum Representation of Multi-Channel Audio), proposed by \cite{extra_ref5:qrma} extends the logic of a state-oriented representation for quantum audio. In our naming system, it is called Multichannel Quantum State Modulation (MQSM). It furthers the QSM implementation logic by introducing a new quantum register in the circuit: the \textit{channel register} (Eq. \ref{eq:mqsm_channelreg}). With this new register, we can address, for example, left ($\ket0$) and right ($\ket1$) channels of a stereo audio (Eq. \ref{eq:mqsm_stereo}). Also, it enables efficient storage of multichannel audio in the same quantum state (or a soundbank multiple mono tracks with the same length). 

\begin{equation}
    (a_{n,m}, t_{n,m}, c_m) \longrightarrow \ket{c_j}\otimes\ket{a_{i,j}}\otimes\ket{t_{i,j}}
    \label{eq:mqsm_channelreg}
\end{equation}

\begin{equation}
\ket{A_{Stereo}}=\frac{1}{\sqrt{2N}}\Bigg[ 
\begin{matrix}
\ket{L}\otimes\big(\ket{0}\ket{0} + \ket{0}\ket{1} +\ket{0}\ket{2}+\ket{0}\ket{3}+ ...\big)\\
\ket{R}\otimes\big(\ket{0}\ket{0}+ \ket{0}\ket{1}+\ket{0}\ket{2}+\ket{0}\ket{3}+...\big)
\end{matrix}
\Bigg]
\label{eq:mqsm_stereo}
\end{equation}

\begin{equation}
    \ket{A_{MQSM}} = \frac{1}{\sqrt{NC}}\sum_{i=0}^{N -1}\sum_{j=0}^{C-1} \ket{c_j}\otimes\ket{a_{i,j}}\otimes\ket{t_{i,j}}
    \label{eq:mqsm}
\end{equation}

\medskip
The general equation of the MQSM is shown in Equation \ref{eq:mqsm}, where $C=2^c$ is the number of channels used. Similar to $N$, if the number of channels is not a power of two, the remaining possible channels would be zero-padded before preparing the state.

\medskip

Correspondingly, a channel register could be added to Coefficient-Based representations. As exemplified in the next section, it can become a valuable tool for feature extraction algorithms. For instance, A Multichannel version of the SQPAM could be derived as shown in Equation \ref{eq:msq-pam}.

\begin{equation}
    \ket{A_{MSQPAM}}  = \frac{1}{\sqrt{NC}}\sum_{i=0}^{N -1}\sum_{j=0}^{C-1} \ket{c_j}\otimes(\cos{\theta_i}\ket{0} + \sin{\theta_i}\ket{1})\otimes\ket{i} 
    \label{eq:msq-pam}
\end{equation}

\section{Summary}
\label{sec:summary}

Different strategies for encoding audio in quantum systems were developed from quantum image representations. A change in the quantum audio representation terminology was also proposed (which could be extended to images and other signals in the future). The key pieces of information and features discussed are compared in Table \ref{table:QAR}.

\medskip
When progressing down the table, the representations require more and more qubit space to store all of the information. As a result, the probabilistic representations are capable of storing larger samples in the same hardware. This is also true for bit depth: since the probability amplitudes are continuous variables, they can store as much amplitude resolution as classical computers can provide after the retrieval process. In comparison, the state-based representations use qubit binary words, which require significantly more quantum space to achieve a comparable resolution. Also, the bit depth $q$ has a role in the preparation complexity of the state modulation.

\medskip
On the other hand, the amount of instructions necessary to prepare the SQPAM state considering only the audio size $N$ is significantly worse than the deterministic representations for larger audio samples.

\renewcommand{\arraystretch}{1.5}
\begin{table}[ht]% Minipage has width exactly the same as the text block
\centering% Centers the contents of the minipage
\scriptsize
\begin{tabular}{p{0.129\linewidth}|p{0.08\linewidth}|p{0.125\linewidth}|p{0.150\linewidth}|p{0.135\linewidth}|p{0.115\linewidth}|p{0.09\linewidth}}
    \toprule
    \textbf{Representation} & \textbf{Original Acronym} & \textbf{Basic Structure} & \textbf{Amplitude Mapping} & \textbf{Space Required} & \textbf{Preparation Complexity} & \textbf{Retrieval} 
    \\ \midrule
    QPAM & ------ & $\alpha_n\ket{t_n}$ & Normalized, strictly positive vector & $\ceil[\big]{\log N}$ & $\order{N}$ & Probabilistic\\
    SQPAM & ------ & $(\cos\theta_n\ket0+\sin\theta_n\ket1)\ket{t_n}$ & Angle Vector & $\ceil[\big]{\log N} + 1$ & $\order{N^2}$ & Probabilistic\\
    uQSM & QRDA & $\ket{U_n}\ket{t_n}$ & Unsigned Integer Vector & $\ceil[\big]{\log N} + q $ & $\order{qN\ceil[\big]{\log N}}$ & Deterministic\\
    QSM & FRQA & $\ket{S_n}\ket{t_n}$ & Signed Integer Vector & $\ceil[\big]{\log N} + q $ & $\order{qN\ceil[\big]{\log N}}$ & Deterministic\\
    fpQSM & QRDS & $\ket{Fp_n}\ket{t_n}$ & Fixed Point Integer Vector & $\ceil[\big]{\log N} + q $ & $\order{qN\ceil[\big]{\log N}}$ & Deterministic\\
    MQSM & QRMA & $\ket{c_i}\ket{S_n}\ket{t_n}$ & Multiple Signed Integer Vectors & $\ceil[\big]{\log N} + q + \ceil[\big]{\log C}$ & $\order{CqN\ceil[\big]{\log N}}$ & Deterministic\\
    \bottomrule
\end{tabular}

\normalsize
    \caption{Comparison between quantum representations of audio}
    \label{table:QAR}
    
\end{table}

\medskip
These comparisons are essential for imagining near-term applications of music-quality audio on quantum hardware. Imagine that a composer wants to process two different audio samples in a quantum machine for some hypothetical artistic reason. The first sample may be coming from a real-time audio signal processing language, with a block of 64 samples that needs to run on a brand-new quantum audio effect. The second one is stored on a computer and has approximately 1.5 seconds of audio, containing $2^{16} = 65536$ samples that will be fed to a large quantum sound database. Both samples have CD quality: A sample rate of 44,100Hz and a $16$-bit resolution. We will compare three representations: QPAM, SQPAM and QSM.

\medskip
According to Table \ref{table:QAR}, QPAM will require 6 qubits and $\sim$ 64 instructions for storing the first sample; 7 qubits and $\sim$ 4096 instructions for SQPAM. QSM will need to allocate 16 more qubits for achieving the necessary amplitude resolution, with a total of 22 qubits and $\sim$ 6144 instructions.

\medskip
For the second audio, things scale up in the time register. QPAM will use 16 qubits and $\sim$ 65 thousand instructions; SQPAM, 17 qubits and a surprising amount of $\sim$ 4.3 billion instructions. Finally, QSM will need 32 qubits and $\sim$ 16 million instructions.

\medskip
A noticeable result of this comparison is that SQPAM might be better suited for processing small signal samples or segmented blocks of audio, useful for algorithms for information retrieval). QSM, on the other hand, would handle large sounds effectively. Also, QPAM needs the lowest amount of resources for preparation, which is very useful for near-term applications.

\subsection{Running on Real Hardware}
\label{subsec:realhardware}
This chapter focuses on surveying and consolidating a smooth and didactic theoretical transition path between digital audio and the new field of quantum audio, underlying potential problems of the computational level, proposing a more unified terminology, and finally glimpsing at potential artistic applications. This approach is why the chapter focuses on a high-level discussion, assuming that the qubits are perfect and perfectly manipulable, referred to as a \textit{logical qubit}. Even so, it is worth saying something about running these algorithms on real hardware. The state-of-the-art computers that are publicly available at the time of writing have a range of three to sixteen rather noisy \textit{physical qubits}. It is possible to correct potential errors for a better approximation of a logical qubit. For that purpose, multiple physical qubits can be combined into a \textit{fault-tolerant} qubit. This field is known as Quantum Error Correction \cite{extra_ref6:qerrc}. The state-of-the-art machines and their qubits are not fault-tolerant.
 
\medskip

A real-world qubit is prone to a myriad of errors and noises. For example, its state can be flipped spontaneously due to external electromagnetic interference. It also tends to lose energy and go to a lower energy level state. Furthermore, a quantum system might lose coherence as time evolves. Also, each gate or instruction applied to a qubit has a chance of failing or being inaccurate, introducing more noise and potentially leading to incorrigible errors. Additionally, qubits might interfere with each other in different ways, depending on their displacement on the actual quantum chip.

\medskip
Unfortunately, the depth of the circuits necessary to prepare quantum audio representations might be too high for near-term, fault intolerant machines (as previously shown in the last section). 
The more instructions present in a fault-intolerant machine (and the more physical qubits you have), the less reliable the result will be\footnote{While this limitation may be true, it can be seen as an advantage inside a sensible noise/degraded aesthetic for artistic purposes.}. This technological barrier has propelled the quantum audio community towards theory and simulations rather than real-hardware experiments on state-of-the-art technology.

\medskip

As a reference for future analysis and comparison, this text underlines how some of the presented encoding schemes (QPAM, SQPAM, QSM) perform in state-of-the-art hardware. An experiment was made using IBM Q hardware \cite{extra_ref7:ibmq}. Specifically, a 7-qubit (32 Quantum Volume) chip called \textit{ibmq\_{casablanca}}. These experiments (along with results and simulations in the following sections) were run by using quantum audio modules implemented in Python and Qiskit, within the scope of the QuTune Project \cite{extra_ref8:qutune}.
The QuTune module was used to map and generate the quantum audio circuits from a signal (shown in Fig. \ref{subfig:originalRH}). Then, the circuit was fed to the IBMQ system. Finally, the resulting histograms - ehxhibited in Figures \ref{subfig:qpamRetrRH}, \ref{subfig:sq-pamRetrRH}, \ref{subfig:qsmRetrRH} - were retrieved and post-processed. 
The reconstructed audio signals are shown in Figures \ref{subfig:qpamRecRH}, \ref{subfig:sq-pamRecRH} and \ref{subfig:qsmRecRH}.

\begin{figure}[h]
\centering
    \includegraphics[width=150pt]{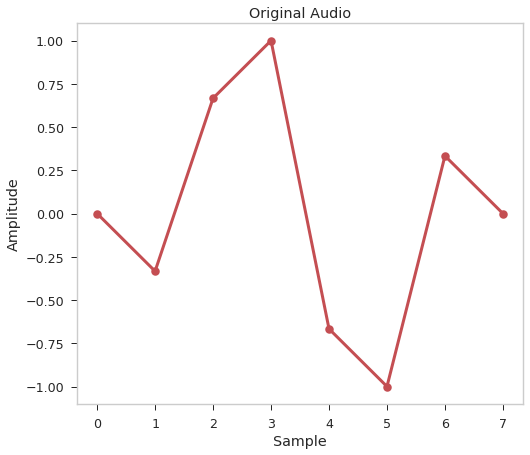}
    \caption{Test audio with 8 samples.}
    \label{subfig:originalRH}
\end{figure}

\medskip
QPAM (Figs. \ref{subfig:qpamRecRH}, \ref{subfig:qpamRetrRH})) had the lowest preparation complexity (approximately 8 instructions and 3 qubits). It seems to retain most of the original information. 

\begin{figure}[ht]
    \centering
    \includegraphics[width=200pt]{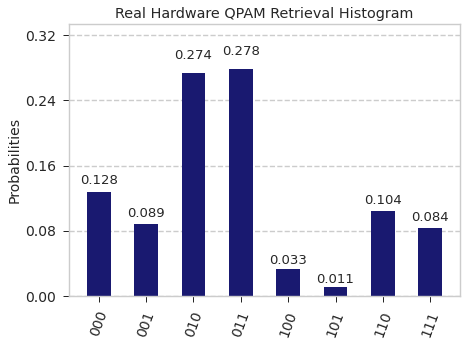}
    \caption{Retrieved QPAM histogram.}
    \label{subfig:qpamRetrRH}
\end{figure}
\begin{figure}[ht]
    \centering
    \includegraphics[width=150pt]{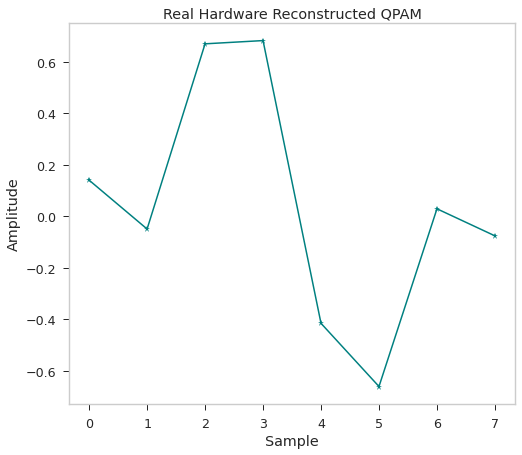}
    \caption{Reconstructed audio from Fig. \ref{subfig:qpamRetrRH}.}
    \label{subfig:qpamRecRH}
\end{figure}

\medskip

The SQPAM (\ref{subfig:sq-pamRetrRH}, \ref{subfig:sq-pamRecRH}), on the other hand, introduced value-setting operations, which are more expensive. The retrieved audio lost most of its amplitude range (notice the values in the y axis) but retained some of the original profile. 

\begin{figure}[ht]
    \centering
    \includegraphics[width=200pt]{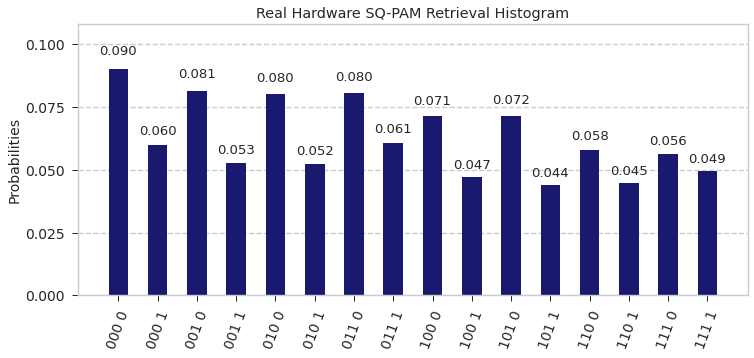}
    \caption{Retrieved SQPAM histogram.}
    \label{subfig:sq-pamRetrRH}
\end{figure}

\begin{figure}[ht]
    \centering
    \includegraphics[width=150pt]{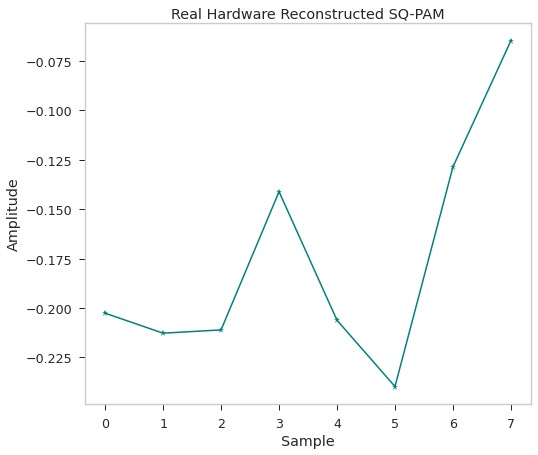}
    \caption{Reconstructed audio from Fig. \ref{subfig:sq-pamRetrRH}.}
    \label{subfig:sq-pamRecRH}
\end{figure}

\begin{figure}[ht]
    \includegraphics[width=\textwidth]{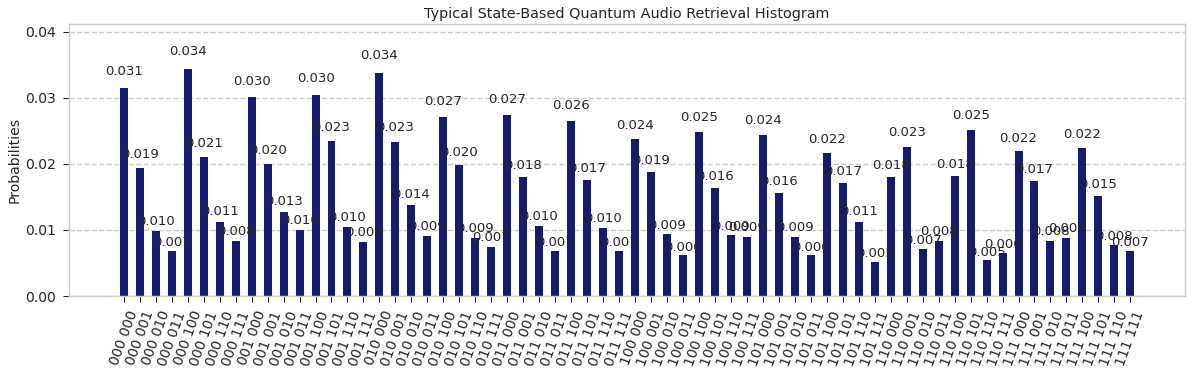}
    \caption{Retrieved QSM histogram.}
    \label{subfig:qsmRetrRH}
\end{figure}
\begin{figure}[ht]
    \centering
    \includegraphics[width=150pt]{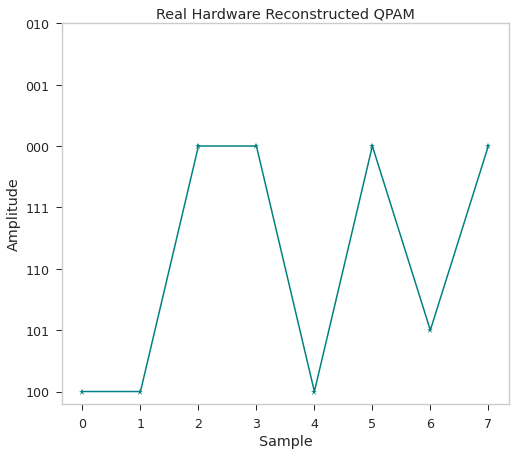}
    \caption{Reconstruction attempt from Fig. \ref{subfig:qsmRetrRH}.}
    \label{subfig:qsmRecRH}
\end{figure}

\medskip

In contrast, QSM (Fig. \ref{subfig:qsmRetrRH}) produced an result. Surprisingly, its resemblance with the original file was gone. This was probably caused by the amount of multi-controlled instructions for preparation and the time required for the operations. Moreover, some qubits might have flipped to a lower energy state. Even though it has a guarantee of exact retrieval, the reconstruction process relies on the states that are most likely to be measured. At this stage of loss of information, the QSM audio was irrecoverable (\ref{subfig:qsmRecRH})\footnote{Still, in the QSM histogram (\ref{subfig:qsmRetrRH}), there is a slight emergence of periodicity on the qubit sequencing. Can it be used artistically? }.

\section{Final Remarks}

This chapter discussed various different approaches for quantum representation of audio that have been proposed to date. Quantum audio is still a very young field of study. Naturally, the different approaches presented here have advantages and disadvantages. At this stage, we are not in a position to foresee which of the method(s) discussed here, if any, might end up being adopted by the industry simply because the hardware to realize them is not available. 

\medskip

Nevertheless, as shown above, we can already start developing approaches and dream of interesting applications (e.g., a quantum electric guitar effects pedal) based on simulations and educated guesses about how quantum processors will evolve. An important caveat of quantum computing that needs to be addressed is the problem of memory. Essentially, quantum information cannot be stored and retrieved in the same way that digital information can with classical processors. The development of quantum random access memory (qRAM) technology would make a significant impact on quantum audio. A number of propositions for developing qRAM have been put forward (e.g., \cite{chen2021, Asaka2021}), but a practical device for more general use is still not in sight yet.

\medskip
{\fontfamily{qcr}\selectfont{quantumaudio}}: A Python package for quantum representations of audio in Qiskit is available: \\ \\
\url{https://pypi.org/project/quantumaudio/} \\

{\fontfamily{qcr}\selectfont{quantumaudio}} is developed and maintained by the quantum computer music team at the Interdisciplinary Centre for Computer Music Research (ICCMR), University of Plymouth, UK.

\section{Acknowledgements}

The authors acknowledge the support of the University of Plymouth and the QuTune Project, funded by the UK Quantum Technology Hub in Computing and Simulation.

\end{document}